\documentclass[lettersize,journal]{IEEEtran}
\IEEEoverridecommandlockouts
\usepackage{cite}
\usepackage{authblk}

\usepackage{amsmath,amssymb,amsfonts}
\usepackage{algorithmic}
\usepackage{algorithm}

\usepackage{graphicx}
\usepackage{textcomp}
\usepackage{fancyhdr}
\usepackage{amsfonts}
\usepackage{xcolor}
\usepackage{amsmath,amssymb,amsfonts}

\usepackage{caption}
\usepackage{subcaption}
\def\BibTeX{{\rm B\kern-.05em{\sc i\kern-.025em b}\kern-.08em
    T\kern-.1667em\lower.7ex\hbox{E}\kern-.125emX}}
\usepackage{times}  
\usepackage[hidelinks]{hyperref}
\usepackage{titlesec}

\usepackage[flushleft]{threeparttable} % threeparttable
\usepackage{multirow}
\usepackage{amsmath,amssymb,amsfonts}
\usepackage{threeparttable}
\usepackage{array}
\usepackage{diagbox}
\usepackage{booktabs}
\usepackage{caption} 
\usepackage{color}
\usepackage{bm,color}
\usepackage{enumitem,kantlipsum}

\graphicspath{{./figures/}}
\newcommand{\name}{SMART}

\begin{document}

\title{A Deep Reinforcement Learning-Based Resource Scheduler for Massive MIMO Networks}

\author{Qing An, Santiago Segarra, Chris Dick,  Ashutosh Sabharwal, Rahman Doost-Mohammady
\thanks{Qing An, Santiago Segarra, Ashutosh Sabharwal and Rahman Doost-Mohammady are with the Department of Electrical and Computing Engineering, Rice University, Houston, TX, USA. E-mail:\{qa4, segarra, ashu, doost\} @rice.edu, Chris Dick is with NVIDIA Corporation, Santa Clara, CA, USA. E-mail: cdick@nvidia.com }% <-this % stops a space
\thanks{This work was supported by the U.S. National Science Foundation
under Grants CNS-1827940, CNS-2016727, and CNS-2120447}
}

% The paper headers
% \markboth{Journal of \LaTeX\ Class Files,~Vol.~14, No.~8, August~2021{Get rid of this prior to submission or put a generic placeholder!!!}}%
% {Shell \MakeLowercase{\textit{et al.}}: A Sample Article Using IEEEtran.cls for IEEE Journals}

% \IEEEpubid{0000--0000/00\$00.00~\copyright~2021 IEEE}

\maketitle
\thispagestyle{fancy}        
\fancyhead{}                     
\lhead{This work is accepted by IEEE Transactions on Machine Learning in Communications and Networking (TMLCN). \\
Copyright is owned by IEEE.}     
\chead{}
\rhead{}
\lfoot{}
\cfoot{\quad}  %current page number
\rfoot{}
\renewcommand{\headrulewidth}{0pt}  
\renewcommand{\footrulewidth}{0pt}
\pagestyle{empty}

%%%%%% -- PAPER CONTENT STARTS-- %%%%%%%%
%%%%%%%%% ABSTRACT
\begin{abstract}
The large number of antennas in massive MIMO systems allows the base station to communicate with multiple users at the same time and frequency resource with multi-user beamforming. However, highly correlated user channels could drastically impede the spectral efficiency that multi-user beamforming can achieve. As such, it is critical for the base station to schedule a suitable group of users in each time and frequency resource block to achieve maximum spectral efficiency while adhering to fairness constraints among the users. 
%\textcolor{red}{Optimal resource scheduling in massive MIMO networks is an NP-hard problem}, with complexity growing exponentially with the number of users.
In this paper, we consider the resource scheduling problem for massive MIMO systems with its optimal solution known to be NP-hard. Inspired by recent achievements in deep reinforcement learning (DRL) to solve problems with large action sets, we propose \name{}, a dynamic scheduler for massive MIMO based on the state-of-the-art Soft Actor-Critic (SAC) DRL model and the K-Nearest Neighbors (KNN) algorithm. Through comprehensive simulations using realistic massive MIMO channel models as well as real-world datasets from channel measurement experiments, we demonstrate the effectiveness of our proposed model in various channel conditions. Our results show that our proposed model performs very close to the optimal proportionally fair (Opt-PF) scheduler in terms of spectral efficiency and fairness with more than one order of magnitude lower computational complexity in medium network sizes where Opt-PF is computationally feasible. Our results also show the feasibility and high performance of our proposed scheduler in networks with a large number of users and resource blocks. 
%\textcolor{brown}{and comparison with almost all state-of-the-art resource schedulers}
\end{abstract}

%%%%%%%%% INTRODUCTION

\section{Introduction}
\label{sec:intro}

\IEEEPARstart{M}{assive} multiple-input multiple-output (MIMO) is one of the key technologies poised to radically improve the spectral efficiency of the current 5G networks and beyond. Through the use of tens or hundreds of antennas at the base station, it can perform multi-user beamforming to serve tens of users in the same time-frequency resource block (RB). However, scheduling which users to serve simultaneously in each RB plays an important role in achieving the large throughput gains promised by the massive MIMO technology. 
%Essentially, in networks with a large number of users, if the wireless channels between two or more of the selected users are largely correlated, the beamforming performance will be poor. 
Beamforming performance can be significantly degraded if there is a substantial correlation in the wireless channels among the scheduled users, as this correlation makes it challenging to effectively focus signal energy when transmitting toward scheduled users. Similarly, separating the signals received from multiple users becomes challenging when their channels are correlated.
%On the other hand, optimally selecting the users in each RB to maximize the overall network throughput and fairness will become increasingly more complex with the number of users and their mobility. Thus, user selection is considered a great challenge in the massive MIMO regime. 
In networks with high user mobility, the channels of individual users and their correlations with other users within each RB are rapidly fluctuating. This dynamic nature of channel characteristics substantially increases the challenges associated with achieving optimal resource scheduling for massive MIMO networks.
%\textcolor{red}{talk about other important factors, such as modulation and coding level selection.}
% \textcolor{red}{What solutions do exist for this problem? \\
%  - Talk about traditional algorithm in both SISO and MIMO. Why they are not suitable for for massive MIMO \\
%  - Talk about existing classical schemes and their complexity. \\
%  - What do these solutions lack in solving and what do we do about it?\\
% }
%Radio resource scheduling has been extensively covered in the literature for many years.
%Given the random nature of the wireless channels, radio resources must be carefully allocated to existing users not only to maximize the utility of the radio resources but to also be fair to the users.
Specifically, fair scheduling of radio resources while maximizing spectral efficiency is essential in real deployments.
%Traditionally, users in single antenna regimes are allocated resources based on a proportional fairness scheme where the selection of the users is weighted by their instantaneous rates divided by the total throughput they have received (fairness criteria).
%\textcolor{red}{Give specific examples.}
%In the multi-user MIMO (MU-MIMO) regime, where multiple users are scheduled in the same RB, this complexity grows even larger, given that selecting multiple users is combinatorial in nature~\cite{castaneda2017overview}.
%\textcolor{red}{
The formulation of the optimal \emph{Proportionally Fair} (Opt-PF) scheduling problem typically results in an integer linear optimization (ILP) problem with an NP-hard solution~\cite{li2008pf}.
%} 
%Strong channel dependency between the users will adversely affect their achievable throughput. Particularly, the instantaneous rate for each user depends on how strongly that user is correlated with other users it is scheduled with. 
The large complexity associated with solving an ILP, when the number of users and resource blocks is large, prohibits designing optimal yet computationally feasible schedulers that can work in the time-stringent 5G and beyond standards. %By optimality, we mean achieving both optimal spectral efficiency and fairness. 
There is a large body of work~\cite{huang2013tvt,Ko2012tc,prasad2014tmc,chen2020twc} that design heuristics or approximation algorithms with low complexity to optimize the spectral efficiency of the networks. However, they either do not evaluate fairness at all or demonstrate poor fairness. This is due to the fact that designing low-complexity approximation algorithms for multi-objective combinatorial optimization problems is typically hard~\cite{chassein2020euor}. %\textcolor{red}{Let's concede the point that this problem is not new and there is a large body of good work, but that the use of deep learning is relatively new. The strength of our work is joint consideration of computational complexity, spectral efficiency, and fairness in mobile environments. The latter is particularly missing in many of the heuristics-based methods where they try to achieve high throughput with low computational complexity. We need to clearly convey this message in the introduction and motivation.}

% There is a recent trend to use deep reinforcement learning (DRL) models to gradually learn from past experience and to learn the best scheduling policy in multi-objective settings~\cite{luong2019cst}. 
% DRL has achieved tremendous success in solving a variety of complex decision-making problems in the past few years. Applications such as robotics control and cyber-physical systems, have benefited from DRL in performing complex tasks without any human-in-the-loop~\cite{mnih2015humanlevel}.
% \IEEEpubidadjcol

% The training of DRL algorithms is guided by a reward function that influences which actions they will take in the future. 
% The goal is to learn the set of actions or policies that will provide the system with the highest reward in any given state. 

% Instead of using an explicit mathematical model, decision optimization in a wireless resource scheduler can be represented as a Markov Decision Process (MDP) whose observations and actions are guided by a well-defined reward function. A DRL agent can then approach an optimum MDP solution by learning from its interactions with the wireless environment. 

In the field of artificial intelligence and machine learning, Markov Decision Processes (MDPs)~\cite{mdp} have emerged as a powerful mathematical framework for modeling decision-making problems under uncertainty. MDPs represent sequential decision processes as a set of states, actions, and transition probabilities, where the goal is to find an optimal policy that maximizes a predefined objective function, such as expected cumulative rewards. However, solving MDPs can be computationally demanding, especially for complex problems with large state and action spaces. To address this challenge, Deep Reinforcement Learning (DRL)~\cite{mnih2015humanlevel} has gained significant attention in recent years. DRL combines reinforcement learning algorithms with deep neural networks to approximate value functions or policies, enabling the handling of high-dimensional state spaces. By leveraging the representation power of deep neural networks, DRL algorithms have achieved remarkable successes in solving continuous and discrete action space problems in various domains, including robotics~\cite{dargazany2021drl}, game playing~\cite{mnih2013playing}, and energy management~\cite{lissa2021deep}. Notably, DRL has also been applied to solve complex combinatorial optimization tasks. For instance,~\cite{bello2016neural} has adopted DRL to solve the traveling salesman problem, a classic combinatorial optimization problem. Similarly,~\cite{li2022csp} solves the covering salesman problem through a DRL model. This motivates the need to explore DRL as a potential tool to solve the optimal proportionally fair resource scheduling for massive MIMO networks.
%The objective is to find the shortest possible route that a salesman can take to visit a set of cities exactly once and return to the starting city. DRL model takes current city as input, outputs the next city and continuous until a complete tour is obtained. \cite{jiang2017deep} utilize DRL to tackle portfolio management, an optimization problem in finance. The goal is to allocate investment funds among different assets to maximize long-term returns while managing risks. The DRL model maps historical asset prices and market indicators to an optimal asset allocation strategy to achieve good portfolio performance. Both these two work show DRL potential for addressing challenging real-world decision-making optimization problems efficiently and effectively. 
%Therefore, DRL also becomes a viable solution for user scheduling in massive MIMO networks which can also be formulated as a MDP model.
%}
Instead of using an explicit mathematical model, decision optimization in a wireless resource scheduler can be represented as a Markov Decision Process (MDP) whose observations and actions are guided by a well-defined reward function. A DRL agent can then approach an optimum MDP solution by learning from its interactions with the wireless environment. 
%\textcolor{red}{ A few recent proposed DRL models for discrete action spaces are suitable for resource scheduler design, such as Deep Q-Network (DQN)~\cite{mnih2013playing}, Double DQN~\cite{van2016deep}, Advantage Actor-Critic (A2C), Asynchronous Advantage Actor-Critic (A3C)~\cite{a3c}, Actor-Critic with Experience Replay (ACER)~\cite{acer}, and Proximal Policy Optimization (PPO)~\cite{ppo}. DQN a landmark algorithm in the field of deep reinforcement learning, which utilizes a Q-network to estimate action values. Double DQN is an extension of the DQN that mitigates overestimation issues by employing two separate neural networks to do action selection and evaluation. A2C and A3C employ actor-critic methods to learn both policy and value functions, with A3C using asynchronous agents to improve efficiency. ACER incorporates experience replay into the actor-critic framework for sample-efficiency and stability, while PPO optimizes policies by iteratively updating them within a trust region. %}
The choice of the DRL model to solve the resource scheduling problem is crucial in achieving high performance and scalability in terms of the number of users in real-world massive MIMO networks. In the recent years, many DRL models for decision making in discrete action space that fit the resource scheduling problem have been proposed. Deep Q-Network (DQN)~\cite{mnih2013playing}, Double DQN~\cite{van2016deep}, Advantage Actor-Critic (A2C), Asynchronous Advantage Actor-Critic (A3C)~\cite{a3c}, Actor-Critic with Experience Replay (ACER)~\cite{acer}, and Proximal Policy Optimization (PPO)~\cite{ppo} are a few examples. However, all these models are shown to struggle with large discrete action spaces that are typically present in combinatorial optimization problems, a phenomenon known as action dimensional disaster~\cite{chen2021eusipco}.
% A few recent works rely on DRL methods such as Deep Deterministic Policy Gradient (DDPG) and Deep Q-learning (DQN) for scheduler design~\cite{huang2021joint,gu2020knowledge,kumar_deepqlearning_2021}. 
% DQN is a discrete-based DRL model used to solve problems with discrete actions. 
%\textcolor{red}{
%Nevertheless, for discrete-control-based DRL models, the large action sets of combinatorial optimization problems (e.g., all user combinations in a larger network) lead to severe convergence issues, which is known as action dimensional disaster~\cite{chen2021eusipco}. Consequently, discrete-control-based model is not directly applicable to large-scale networks with tens or hundreds of active users and discretizing continuous-control-based is a viable solution.
Another class of DRL models that deal with continuous action spaces has been used and adapted for discrete action spaces in various domains.
For instance, Deep Deterministic Policy Gradient (DDPG)~\cite{ddpg} is a popular continuous-based DRL model used to solve a variety of decision problems with large discrete action spaces~\cite{dulac2015drl}, including resource scheduling in massive MIMO~\cite{guo2020globecom,chen2021joint}. However, DDPG is known to be very sensitive to hyper-parameter tuning in actual training, especially in high-dimensional and complicated tasks~\cite{sac2019}.
%which combines deep neural networks and deterministic policy gradients to optimize continuous action spaces. 
%Trust Region Policy Optimization (TRPO)~\cite{trpo} focuses on stable policy updates by enforcing a constraint on policy changes, ensuring a conservative yet effective exploration of the policy space. Additionally, both A3C and PPO are able to deal with problems with continuous action space.
%However, the required scalability is far below what is required for user scheduling in real-world massive MIMO networks because of non-enough sample-efficiency and scalability~\cite{sac2018}. Furthermore, most of state-of-the-art DRL models are very sensitive to hyper-parameters tuning in actual training, especially in high-dimensional and complicated tasks~\cite{sac2019}.
%} 

In this paper, we present a novel DRL framework for the resource scheduling problem in massive MIMO networks. 
The novelty of our framework is three-fold:

First, we propose a DRL-based scheduler design named \name{}, based on the recently proposed soft actor-critic (SAC) model~\cite{sac2018}. 
The SAC model has superior sample efficiency by incorporating an entropy term in its value function and automatic tuning of hyper-parameters. 
Therefore, it can converge to the optimal solution in large multi-dimensional action spaces much faster than the existing models such as DDPG. Given that SAC is by design used for continuous space problems, we propose to combine SAC with K-Nearest Neighbors (KNN) algorithm to generate discrete outputs corresponding to user scheduling decisions in massive MIMO networks. To achieve the scalability required for real-world massive MIMO networks with a large number of users, we propose a novel dimension division strategy that maps the discrete action set for scheduling to multiple dimensions. 
% To the best of our knowledge, our work is the first application of a discretized continuous-based DRL model to the wireless resource scheduling problem.

Second, we significantly reduce the state space and, thus, the complexity of the proposed SMART model for massive MIMO by using user grouping labels as the model states instead of the raw channel state information (CSI) matrix. 
The user grouping labels indicate which users have less correlated channel vectors, hence, are more suitable to be scheduled at the same time. 
This reduces the computational complexity of the model in both training and inference by $\mathrm{2}\times$ without sacrificing spectral efficiency or fairness.
% Additionally, since the SAC model is originally designed for continuous action spaces, we adapt the SAC to work in very large discrete action spaces as required in large networks.

Third, we demonstrate the scalability of \name{} to a large number of resource blocks consistent with 5G systems. 
We demonstrate that our scheduler framework can operate independently on different resource blocks and, at the same time, achieve close to optimal performance.

%\textcolor{teal}{\textbf{Multiple Resource Blocks.} We first implement resource scheduler for one resource block and then extend it to multiple resource blocks, which makes it more adaptable to 5G technology in which there are tens or hundreds of resource blocks to be assigned among users in a very stringent time slot.} 

We evaluate the effectiveness of \name{} in various channel conditions in both simulated as well as real-world channel traces
through a comparison of its performance with state-of-the-art scheduling algorithms, including heuristic-based and DRL-based models. We comprehensively demonstrate the effectiveness of our proposed method in achieving near-optimal spectral efficiency while simultaneously maintaining superior inter-user fairness very close to the Opt-PF scheduler. We experimentally analyze the computational complexity of our method and demonstrate its efficiency. We also provide guidelines on how our proposed system can be deployed on real-world 5G and beyond systems while achieving the latency required for the 5G new radio (NR) standard.
% \textcolor{teal}{Lastly, we also collect some real-world dataset including static and mobility scenarios. To make it comprehensive, we provide numerous cases with varying speeds and inter-user correlation situation. We evaluate effectiveness of our resource scheduler with real-world dataset and demonstrate that the SAC-based scheduler is able to achieve outstanding performance as it does in simulation dataset.}

% \textcolor{red}{Forth: multiple resource blocks. Fifth: Real-world dataset.}
%The remainder of this paper is organized as follows. We present a system model and motivate our work in \S\ref{sec:rw}. We present 
%%%%%%%%% Related work
\section{System Model and Existing Work}
\label{sec:rw}
 \begin{figure}[t]
      \centering
      \includegraphics[width=0.3\textwidth]{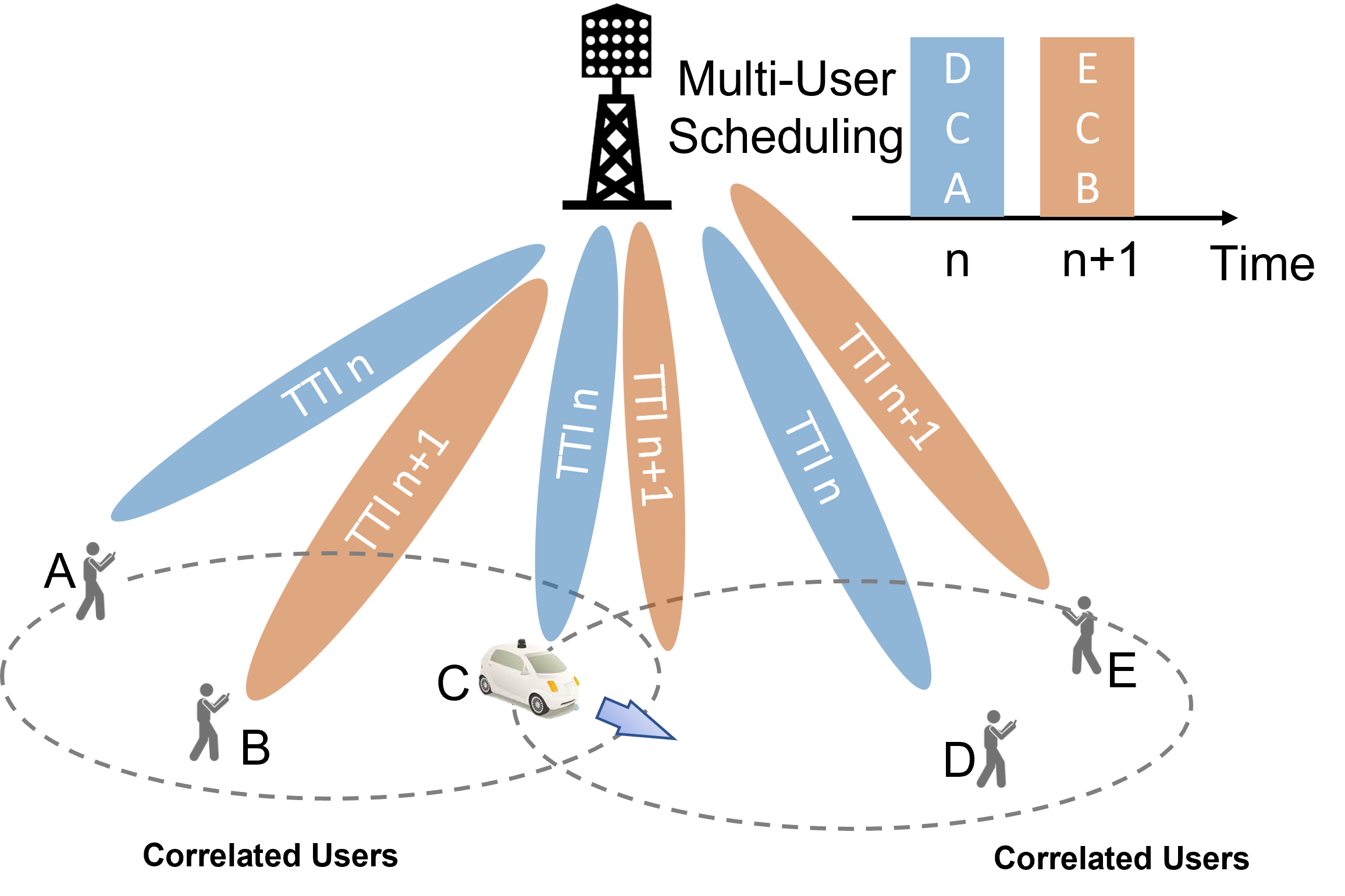}
      \caption{System Model.} 
      \label{fig:system}
\end{figure}

 \subsection{System Model}
\label{sec:system_model}
%\textcolor{red}{Describe the massive MIMO network and add a figure, define the signal model, beamforming method based on CSI, MCS, spectral efficiency.}
We consider a single-cell network with a massive MIMO base station (BS) with $M$ antennas serving $L$ single-antenna users in its cell. 
The base station uses orthogonal frequency division multiplexing (OFDM) and performs MU-MIMO transmission and reception to $N<L$ users such that $N \leq M$.
%In the massive MIMO regime, it is generally assumed that $N \ll M$. 
We consider time-division duplex (TDD) operation, where all $L$ users periodically send orthogonal pilot sequences to the BS for channel estimation. 
We assume that the scheduler possesses full knowledge of the channel condition of all users associated with the BS and the channel for each user does not change during a transmission time interval (TTI). %Even in practical wireless communication systems, partial observability of the channel condition is more common. It is also important to note that partial observability impairs system performance due to outdated channel information, particularly in scenarios involving high-speed mobility. Nevertheless, in this case, we can use complementary methods that perform channel prediction based on the partial channel information such as the ones proposed in~\cite{channelprediction,8949454,8693948}.
% The BS will schedule the periodic transmission of pilots so that it is always updated with the latest channel changes. 
Subsequently, the BS selects a set of $N$ users for data transmission and reception through beamforming based on their estimated channel and assigns their modulation schemes, and communicates that information through the control channel. 
Using their assigned modulation scheme, the selected users will transmit their symbols at the same RB in the uplink and receive them simultaneously in the downlink.
A simplified system model is depicted in Fig.~\ref{fig:system}. 
For the uplink, we consider the following signal model
\begin{equation}
    \textbf{y} = \textbf{H}\textbf{u} + \textbf{n},
\end{equation}
where $\textbf{y}$ is the $M\times 1$ received signal vector at the BS, $\textbf{H}$ is the $M\times N$ channel matrix, and $\textbf{u}$ is the $N\times 1$ transmitted symbols vector by the users. 
Additionally, $\textbf{n}$ is $M\times 1$ receiver complex noise vector with a circular Gaussian distribution, $\textbf{n} \sim C\mathcal{N}(0, \sigma^2\textbf{I})$ where $\sigma^2$ is the noise variance and $\textbf{I}$ is the identity matrix. 
Note that, the value of $N$ can vary in each TTI depending on the current channel condition and it can be bounded by a maximum value $N_{\mathrm{max}}$.
We assume the BS uses zero forcing (ZF) for beamforming. 
The BS calculates the ZF beamformer using the estimated channel $\hat{\textbf{H}}$ as
\begin{equation}
    \textbf{W} = \hat{\textbf{H}}(\hat{\textbf{H}}^H\hat{\textbf{H}})^{-1}.
\end{equation}

The BS then performs receive beamforming on the received signal to estimate the transmit symbol vector $\hat{\textbf{x}}$ as

\begin{equation}
    \hat{\textbf{u}} = \textbf{W}^H\textbf{y}.
\end{equation}

For simplicity, we only consider the uplink, but the above model is extendable to the downlink as well. 
The above signal model is for a single subcarrier in an OFDM system, but the same model applies to all subcarriers. %As we will discuss later, that assume scheduling decisions are made in a resource block (RB) granularity. In 5G-NR, each RB consists of 12 subcarriers. The number of RBs are variable and depends of the total bandwidth.  %Note, we assume the BS allocates all subcarriers of OFDM symbols within each TTI to the selected users. The reason is that in massive MIMO regime, the channel response for each user is smoothed out by the large number of antennas and thus all channel vectors can be considered equally good~\cite{yang2014vtc}.

An RB is the smallest scheduling granularity in 5GNR, which contains resources in the time and frequency domain. One RB in 5G is made up of 12 consecutive subcarriers in the frequency domain~\cite{3gpp38211}. 
In the time domain, the composition of RBs in 5G is more flexible and can vary between one OFDM symbol and the entire slot ($1~\mathrm{ms}$ in numerology 0). The quality of the wireless channel changes dramatically over time, across users, and among different frequency bands. 
%This effect is known as frequency selective fading. 
It is shown in~\cite{chenchannel} that wireless channel capacity might fluctuate by up to 9 times in 20 MHz LTE bandwidth with over 100 RBs. This effect is more pronounced in 5G since it typically has a wider bandwidth (i.e. 40 MHz to 400 MHz). Consequently, user selection decisions will vary across RBs due to the frequency selectivity of the channel. Thus, it is essential to take into account resource scheduling for every RB individually. In our design, we first focus on resource scheduler design on a single RB and then extend to many RBs to show the adaptability of our proposed scheduler to 5G massive MIMO networks.

\textit{Optimal Schedulers:} In the literature, multiple schedulers are defined as optimal. The rate-optimal scheduler, known as \\emph{Optimal Maximum Rate} (Opt-MR), finds the resource scheduling solution in each TTI that maximizes the sum rate
\begin{align}
\label{eq:max_tp_multi_rb}
      \underset{{x_{l,b}^{t}}}{\mathrm{argmax}} \quad & \sum_{b=1}^{B} \sum_{l=1}^{L} r_{l,b}^{t}\ x_{l,b}^{t},\\
      \mathrm{s.t.} \quad  & \sum_{l=1}^{L} x_{l,b}^{t} \le N_{max} \nonumber \\
      &  x_{l,b}^{t}\in \{ 0,1 \} \nonumber %, \forall l\in \{ 1,2,...L\}, \forall b\in \{ 1,2,...B \}
\end{align}
where $x_{l,b}^{t}$ represents the binary selection of user $l$ at TTI $t$ and RB $b$ and $r_{l,b}^{t}$ is the instantaneous rate achieved by user $l$ at TTI $t$ and RB $b$. We calculate the instantaneous rate as $r_{l,b}^{t} = log_2(1+\mathrm{SINR}_{l,b}^{t})$, where $\mathrm{SINR}_{l,b}^{t}$ is the received signal to interference-plus-noise ratio from each beamformed user $l$ at TTI $t$ and RB $b$. We consider $B$ as the maximum number of RBs being used in the system.

Simply maximizing the sum rate ignores the notion of fairness where, depending on the channel conditions, some users may never get selected. Therefore, a commonly used scheduler, known as \emph{Optimal Proportionally Fair} (Opt-PF) scheduler, finds the resource scheduling solution that maximizes the following objective~\cite{lau2005tc,approx}
% \begin{align}
% \label{eq:pf}
%       \underset{{x_{l}^{t}}}{\mathrm{argmax}} \quad & \sum_{l}^{\mathcal{L}} w_{l}^{t}\ x_{l}^{t},\\
%       \mathrm{s.t.} \quad &  x_{l}^{t}\in \{ 0,1 \}, \forall l\in \{ 1,2,...\mathcal{L} \}\\
%                           &  w_{l}^{t} = \frac{r_{l}^{t}}{p_{l}^{t}}, \forall l\in \{ 1,2,...\mathcal{L} \}\\
%                           &  p_{l}^{t} = 
%                           \begin{cases}
%                               p_{l}^{t-1} + r_{l}^{t-1},\ if\ x_{l}^{t-1} = 1\\
%                               p_{l}^{t-1},\ otherwise
%                           \end{cases}                                                      
% \end{align}
% where $w_{l}^{t}$ denotes the weighted rate, which we calculate it as the ratio of instantaneous rate $r_{l}^{t}$ to received rate $p_{l}^{t}$ until TTI $t$. Normalizing instantaneous rate with the total received rate guarantees that all users have a fair chance of getting selected by the scheduler even when they are experiencing a poor channel.
% If considering multiple RBs, (\ref{eq:pf}) can be rewritten as:
\begin{align}
\label{eq:pf_multi_rb}
      \underset{{x_{l,b}^{t}}}{\mathrm{argmax}} \quad & \sum_{b}^{B} \sum_{l}^{L} w_{l,b}^{t}\ x_{l,b}^{t},\\
      \mathrm{s.t.} \quad & \sum_{l=1}^{L} x_{l,b}^{t} \le N_{max} \nonumber\\
      &  x_{l,b}^{t}\in \{ 0,1 \}\nonumber\\ % , \forall l\in \{ 1,2,...L \}, \forall b\in \{ 1,2,...\mathcal{B} \}
      &  w_{l,b}^{t} = \frac{r_{l,b}^{t}}{\sum_{b}^{B}\ p_{l,b}^{t}}, \label{eq:pf_mid}\\ %\forall l\in \{ 1,2,...\mathcal{L} \}, \forall b\in \{ 1,2,...\mathcal{B} \}
      &  p_{l,b}^{t} = 
      \begin{cases}
          p_{l,b}^{t-1} + r_{l,b}^{t-1},\ if\ x_{l,b}^{t-1} = 1\\
          p_{l,b}^{t-1},\ otherwise
      \end{cases} \nonumber        
\end{align}
where $w_{l,b}^{t}$ denotes the weighted rate, which we calculate as the ratio of instantaneous rate $r_{l,b}^{t}$ to received rate $p_{l,b}^{t}$ until TTI $t$ on all RBs. Normalizing the instantaneous rate with the total received rate guarantees that all users have a fair chance of getting selected by the scheduler even when they are experiencing a poor channel. %\textcolor{red}{In~\eqref{eq:pf_mid}, if we jointly consider the fairness of overall system, we should replace $p_{l,b}^{t}$ with $\sum_{b}^{B}\ p_{l,b}^{t}$.}

%Similarly,~\eqref{eq:pf_multi_rb} can also be formulated as an ILP:
Both optimization problems in~\eqref{eq:max_tp_multi_rb} and~\eqref{eq:pf_multi_rb} are NP-hard since they can be reformulated as an Integer Linear Programming (ILP) problem~\cite{ilp}. Specifically, we can reformat~\eqref{eq:pf_multi_rb} when $B=1$ as the following ILP problem,
\begin{align}
    \label{eq:pf_ilp}
    \underset{\mathbf{x}}{\mathrm{argmax}} \quad & \mathbf{w}^T\ \mathbf{x}\\
    \mathrm{s.t.} \quad & \mathrm{\mathbf{J}}_{\mathrm{L},\mathrm{L}}\mathbf{x} \le N_{max}\mathbf{J}_{\mathrm{L}} \nonumber \\
    &  \mathbf{x}\in \{ 0,1 \}^L \nonumber
\end{align}
where $\textbf{w}$ is a vector of all users instantaneous rates, $\textbf{x}$ is user binary selection vector. Also $\mathrm{\textbf{J}}_{\mathrm{L},\mathrm{L}}$ and $\textbf{J}_{\mathrm{L}}$ are square matrix and vector of all ones with size L, respectively.%\textcolor{cyan}{We can improve the presentation of \eqref{eq:pf_ilp}}
% \begin{equation*}
% \begin{aligned}
%     \textbf{w} = \left[ w_{1}^{t},w_{2}^{t},...w_{L}^{t} \right],\quad 
%     X &= \left[x_{1}^{t}, x_{2}^{t},...x_{L}^{t} \right]
% \end{aligned}
% \end{equation*}

% \begin{equation}
%     \underset{X}{\mathrm{argmax}} \quad W\ X
% \end{equation}
% where $W$ is weighted-rate matrix and $X$ is user selection matrix,
% \begin{equation*}
% \begin{aligned}
%     W = \left[ w_{1}^{t},w_{2}^{t},...w_{\mathcal{L}}^{t} \right],\quad 
%     X &= \begin{bmatrix}
%            x_{1}^{t} \\
%            x_{2}^{t} \\
%            \vdots \\
%            x_{\mathcal{L}}^{t}
%          \end{bmatrix}
% \end{aligned}
% \end{equation*}

%Due to ILP is NP-hard~\cite{ilp}, the \textcolor{red}{Opt-PF and Opt-MR} are NP-hard.
Solving \eqref{eq:pf_ilp} by exhaustively searching through the combinations of vector $\textbf{x}$ has the complexity of $\mathcal{O}$($2^{L}$). Solving \eqref{eq:max_tp_multi_rb} and \eqref{eq:pf_multi_rb} through an exhaustive search, when $B$ RBs are considered, the complexity will increase to $\mathcal{O}$($2^{LB}$). 
%The complexity exponentially increases as the number of users grows up and becomes prohibitive in large-scale massive MIMO networks. 
However, there are approximate algorithms for the Opt-PF problem with polynomial complexity, such as the one proposed in~\cite{approx}. We discuss and evaluate an approximate algorithm in~\S\ref{sec:exp} along with other benchmarks.

\subsection{Existing Work and Motivation}
\label{sec:existingwork}
Recent work on resource scheduling in massive MIMO and MU-MIMO can be classified into two general categories: heuristics schedulers, and AI-based schedulers. In this section, we provide an overview of some of the most relevant works in each category.
%Resource scheduling for MU-MIMO, including user scheduling and MCS selection methods, has been considered in recent work (e.g., ~\cite{chen2021mcore,yang2018spawc,ddpg,shekhawat2020globecom,wang2021pimrc,mauricio2020icc}). Most research work focus on MU-MIMO ~\cite{chen2021mcore,wang2021pimrc,mauricio2020icc,shekhawat2020globecom} and they can be classified into two categories, optimization work and Reinforcement Learning (RL)-related work, based on their methods adopted. But there are still some work ~\cite{ddpg,yang2018spawc} about massive MIMO scheduling. This task is more challenging.  

\textit{Heuristics Scheduler Designs}:
Many existing MU-MIMO scheduling works provide heuristics-based approximations to the Opt-PF scheduler~\cite{chen2020twc,liu2017systems,chen2021mcore}. While they try to strike a balance between complexity and performance, often their complexity does not scale to large networks or they significantly underperform the optimal scheduling policies.

%\textcolor{red}{Briefly talk about each. Be more specific about their pros and cons and try to motivate AI-based along the way.}
The scheduler proposed in~\cite{chen2021mcore} implements a multi-phase optimization to solve Eq.~(\ref{eq:pf_multi_rb}) in MU-MIMO settings. 
It narrows down the exhaustive search needed for the Opt-PF solution using some relaxations of the optimization problem. For e.g., it decouples the user selection in different RBs. Moreover, in each RB, it reduces the number of choices based on the channel quality of each user before deciding the user selection action based on the correlation of the remaining users. % \textcolor{red}{maybe add a few words about what are these `some heuristics'?}. 
Through these sub-optimal relaxations, their method can be parallelized and efficiently implemented on a powerful GPU, and hence can meet the stringent 5G-NR latency constraints (i.e., nearly $1 \mathrm{ms}$). Despite the low-latency implementation,, this scheduler only scales to $M=12$ and $N=4$, and as a result, it has limited scalability to massive MIMO.
In~\cite{chen2020twc}, two heuristics-based user scheduling algorithms are proposed and evaluated on channel datasets collected from a dense indoor massive MIMO network with stationary users. However, the algorithms sacrifice fairness in favor of spectral efficiency. They are also not evaluated under mobility scenarios.
%In~\cite{chen2020twc}, two heuristics-based user scheduling algorithms which can keep inter-user fairness while alleviating correlations and pilot overhead are proposed and evaluated on channel datasets collected from a dense indoor massive MIMO network with stationary users. However, each time it makes a decision, scheduler needs to calculate all inter-user correlation between selected users and users not chosen. It's a greedy search which sacrifices computation complexity to improve spectral efficiency. Furthermore, this algorithm is designed for predefined scheduled user number and lacks the comparison between cases with different scheduled user number. It is also not clear how they perform under mobility scenarios.
The work in~\cite{yang2018spawc} proposed a scheduler for massive MIMO that schedules users with low correlation channels in the same time slot. It first partitions users into groups through a user grouping algorithm. The scheduler then goes through all groups and schedules all users in each group with a rate-fair method. As we discuss later in the paper, this scheduling algorithm fails to work well in fast-varying channel environments when inter-user channel correlations are continuously changing and it is unable to fairly allocate users across channel coherence blocks.

\textit{AI-based Scheduler Designs}: 
Due to the huge complexity of the optimization-based methods, several recent works~\cite{guo2020globecom,chen2021eusipco,shi2018wcsp,bu2019iccc,chen2021joint,lopes2022deep,10038693} have proposed DRL models for MIMO scheduling. %\textcolor{blue}{Finish: Talk about each and mention the most relevant ones and the type of DRL model they use.} 
A Q-learning-based DRL resource scheduling is proposed in~\cite{bu2019iccc}. 
It models the user scheduling problem as a Markov Decision Process (MDP) that outperforms the round-robin scheduler in terms of sum rate. 
However, the discrete DRL models are known to have difficulty in converging in large action sets~\cite{zhao2018icaidb}. 
The convergence issue is also true for more advanced discrete DRL models, such as DQN and Double DQN.
As such, discrete DRL models have limited scalability to a large number of users for multi-user scheduling in massive MIMO networks. We will also demonstrate these limitations in~\S\ref{sec:exp}.

The work in~\cite{guo2020globecom} proposes a DDPG-based user scheduler for massive MIMO networks. 
Its model outputs a probability distribution over all selectable users and chooses the most promising UE combinations at each TTI. 
However, it includes a raw channel matrix in state space and the number of elements in action space equals the number of UEs. Large state and action spaces hinder its scalability. 
This algorithm is extended in~\cite{chen2021joint} for both user scheduling and transmit precoding based on DDPG. 
It considers multiple antennas and antenna correlation on the UE side as well. However, their proposed scheduler has limited scalability and does not consider the evaluation of user fairness.
%\textcolor{brown}{The method proposed in \cite{chen2021joint} is an improved version of \cite{guo2020globecom} which uses DDPG for user scheduling and transmit precoding by considering multiple antennas and antenna correlation at the user side. Moreover, they use the square of the channel correlation measure matrix as the state elements and reduce it further by half based on the Hermitian features of this matrix. Nevertheless, the number of elements in action space is expanded to the number of antennas on the user side. Large action space still impedes scaling it to a large network.} 
We implement a DDPG-based scheduler as one of our benchmarks and discuss its performance with respect to our proposed scheduler.
A pointer network is investigated in~\cite{chen2021eusipco} as the actor in an actor-critic framework to convert the combinatorial problem in multi-user scheduling into a sequential selection problem. 
However, sequential scheduling has slow inference, which makes it undesirable for latency-sensitive 5G networks. 
Additionally, applying the model to large networks results in a complicated network structure and a long model update time due to the use of a raw channel matrix as the input.
This is exacerbated further by complex-valued channels, which need to be separated into real and imaginary parts before being fed to the model. 
We implement a pointer network-based DRL scheduler as a benchmark and discuss these limitations in more detail in~\S\ref{sec:exp}. 

% The methods proposed in \cite{lopes2022deep,10038693} focus on cross-layer resource scheduling in massive MIMO networks.
% In particular,~\cite{lopes2022deep} utilizes DQN to act as a user scheduler on multiple frequency bands by considering the channel quality indicator (CQI) from the PHY layer and user data traffic demand and packet age from higher layers in the DRL model. Similarly,~\cite{10038693} proposes a DDPG-based algorithm that includes CQI, the requested data size, and the data type of users as elements in the state space.~\textcolor{red}{Although cross-layer design is an intriguing subject that we plan to investigate in future research, we can substitute PHY-measured data rates with cross-layer rates in our work without requiring any modifications. Nonetheless, considering the superior sample-efficiency and scalability of our method, replicating our work can enhance the performance of both studies. We will show the superiority of our proposed method to the algorithms adopted in these work in~\S\ref{sec:exp}.} \textcolor{cyan}{It might be better to remove the cross layer references altogether. We can provide the above explanation in the rebuttal.}
% In our proposed method, which we will discuss next, we do not consider the higher layer parameters, but we use both DQN and DDPG as our benchmarks to show the superiority of our proposed method to these algorithms.

\textit{Our Proposed Method}:
We propose \name{}, a massive MIMO user scheduler based on the recently proposed soft actor-critic (SAC) DRL model~\cite{sac2018,sac2019} and the KNN algorithm~\cite{muja2014knn}. 
SAC has gained attraction in several real-time control problems such as robotic locomotion~\cite{haarnoja2018walk}. 
SAC was originally designed to handle continuous action spaces. However, the user scheduling is a discrete decision problem where an appropriate set of users must be selected at each TTI.
The work in~\cite{christodoulou2019sacdiscrete} provides a modification of SAC for discrete action spaces, but we find that their modification is still not suitable for large discrete action sets as it has serious convergence issues in large-scale networks. 
% \textcolor{blue}{RD: Agree with Santiago. I think we should remove from this sentence up to the end of the paragraph.} Compared to discrete action space, actor can generalize on continuous action space~\cite{dulac2015drl} \textcolor{red}{SS: This discussion seems a bit informal} Consequently, continuous control DRL model is a better choice for solving problems which have large discrete action space. The only thing we need to do is mapping continuous actions to corresponding discrete ones.
Inspired by the approach in~\cite{dulac2015drl}, we use the KNN~\cite{muja2014knn} to discretize SAC to adapt it to discrete action spaces. The basic idea is to use a continuous-based algorithm to generate an initial or ``proto" continuous action first. 
Then, the K nearest discrete actions are found by using the KNN algorithm. 
Among the K nearest discrete actions, the one with the maximum Q value is selected. We further propose a novel dimension division strategy that helps to scale up the size of the combinatorial action set (i.e., number of users in the network) and enhance model convergence capability. Using this approach, we enable our model to dynamically select the users to maximize system spectral efficiency and inter-user fairness. More details are illustrated in~\S\ref{subsec:discrete}. In contrast to prior work, our proposed scheduler is more scalable and performs very close to the Opt-PF solution. 

\section{\name{}: A Scalable SAC-KNN-based massive MIMO Scheduler}
\label{sec:scheduler}
In this section, we first provide a brief introduction to SAC. 
Subsequently, we describe the design of our proposed scheduler based on the SAC DRL framework. We discuss how we discretize the output of the SAC framework by applying the KNN algorithm and propose a dimension division strategy to scale up the supported size of the action set. 
We also propose to reduce the complexity of the framework by using the user grouping instead of the raw channel matrix as the input to the framework. 
Additionally, we discuss how we scale up the model to support as many RBs as needed for realistic 5G networks.
%\textcolor{brown}{We then describe how we innovatively apply our scheduler originated from continuous-based DRL model in the discrete multi-dimensional action spaces to make scheduling decisions. Additionally, user grouping strategy which is used to simplify state space is introduced and scalability of proposed scheduler is also discussed.}
\subsection{A Primer on SAC}
\label{sec:SAC}
%\textcolor{red}{SAC is an off-policy Deep Reinforcement Learning (DRL) model that aims to maximize entropy. To achieve this, SAC employs a stochastic policy, in contrast to the deterministic policy used in Deep Deterministic Policy Gradient (DDPG). Instead of selecting the optimal action, a stochastic policy outputs probabilities for all possible actions. Consequently, the entropy of policy $\pi$ can be obtained using~\eqref{eq:entropy}, while~\eqref{eq:sac} illustrates the optimal policy $\pi^{*}$ for the SAC DRL model, which maximizes both the cumulative reward $R$ and the policy entropy $H$~\cite{sac2018}. By maximizing policy entropy, SAC encourages the model to extensively explore the action space, facilitating the discovery of global optima and enhancing sample-efficiency. Moreover, SAC samples transition from replay memory to learn from past experience, similar to other off-policy algorithms like DQN~\cite{mnih2013playing} and Double DQN~\cite{van2016deep}. In contrast to on-policy models such as PPO~\cite{ppo} and A3C~\cite{a3c}, which update their policies based on experiences generated by the current policy, SAC has the ability to learn from a broader spectrum of experiences. This characteristic enhances sample-efficiency and aids in facilitating convergence.}

%In~\eqref{eq:sac} we show the optimal policy $\pi^{*}$ for the SAC DRL model, which maximizes the cumulative reward $R$ and entropy $H$~\cite{sac2018} 
SAC is an off-policy Deep Reinforcement Learning (DRL) model that employs a stochastic policy, in contrast to the deterministic policy used in Deep Deterministic Policy Gradient (DDPG). Instead of selecting the optimal action, a stochastic policy outputs probabilities for all possible actions. The optimal policy in SAC, defined in ~\eqref{eq:sac}, aims to maximize both the cumulative reward $R$ and the policy entropy $H$.
\begin{equation}
    \pi ^{*} = \arg \max_{\pi } \mathbb{E}_{(s_{t},a_{t})\sim \rho _{\pi }}\left [ \sum_{t}R(s_{t},a_{t}) + \alpha H(\pi(\cdot |s_{t}))\right ].
    \label{eq:sac}
\end{equation}
where the policy entropy $H$ is defined as
\begin{equation}
    H(\pi(\cdot |s_{t})) = -\sum P(a_{t}|s_{t})\times log(P(a_{t}|s_{t}))
    \label{eq:entropy}
\end{equation}

By maximizing policy entropy, SAC encourages the model to extensively explore the action space, facilitating the discovery of global optima and enhancing sample efficiency. Moreover, SAC samples transition from replay memory to learn from past experience, similar to other off-policy algorithms like DQN~\cite{mnih2013playing} and Double DQN~\cite{van2016deep}. In contrast to on-policy models such as PPO~\cite{ppo} and A3C~\cite{a3c}, which update their policies based on experiences generated by the current policy, SAC has the ability to learn from a broader spectrum of experiences. This characteristic enhances sample efficiency and aids in facilitating convergence, especially in high-dimensional action spaces as demonstrated in~\cite{sac2018}.
%The goal of learning is to find an optimal policy that maximizes both the reward and entropy concurrently and, thus, improves sample efficiency. 
 
In general, SAC has the following two major benefits:
\begin{enumerate}[wide, labelwidth=!, labelindent=0pt]
  \item \textbf{Strong exploration capability}. SAC does not discard any action, even if it is not the best one. 
  If multiple promising actions are found, the stochastic policy will choose them with equal probability. 
  This feature helps SAC explore more and not easily get trapped in local optima. 
  In contrast, the deterministic policy-based algorithms, such as DDPG\cite{ddpgalg}, save the action with the highest value resulting in fewer exploration opportunities. 
  
  % \textcolor{red}{Furthermore, the experience replay buffer helps SAC efficiently utilize previous experience to do exploration and learning. This approach demonstrates superior sample efficiency when compared to on-policy DRL models such as PPO~\cite{ppo} and A3C~\cite{a3c}.}\textcolor{cyan}{You already mentioned this above. Remove this.}
  \item \textbf{High robustness}. Most applications of RL require the agent to perform well in the presence of disturbances in the environment. 
  Because of the adopted stochastic and entropy maximizing algorithm, SAC explores as many potential actions as possible and, hence, it is able to deal with complicated and dynamic environments (e.g., mobility scenarios in wireless communication), including scenarios it has never encountered~\cite{eysenbach2021maximum}. 
  %it is able to do adjustment in the face of interference, such as model and estimation errors as stated in~\cite{haarnoja2017reinforcement}. 
  %\textcolor{blue}{SS: It is hard to understand this point. What does interference have to do with this?}
\end{enumerate}

\begin{figure}[t]
      \centering
      \includegraphics[width=0.45\textwidth]{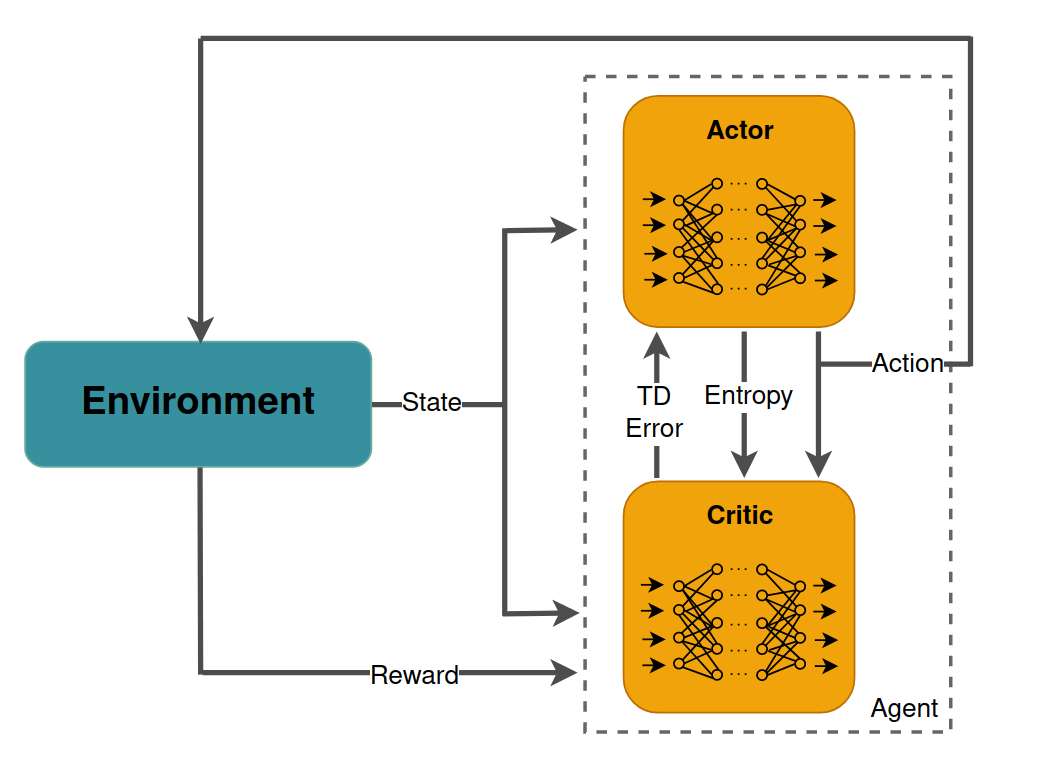}
      \caption{Soft Actor-Critic Framework.} 
      \label{fig:sac}
\end{figure}

Fig.~\ref{fig:sac} shows the block diagram of the SAC framework. 
Similar to any actor-critic architecture in DRL, the actor in SAC generates a policy from which an action is drawn based on the current state. 
The role of the critic is to assess the actor's policy and guide the actor toward the optimal path through feedback. 
Unlike other actor-critic models, SAC adjusts the $Q$ function by a temperature coefficient ($\alpha$ in~\eqref{eq:sac}), which represents the weight of entropy. 
Furthermore, in~\cite{sac2019}, the authors improve SAC with automatic entropy coefficient adjustment. 
This method significantly reduces the burden of manually adjusting hyper-parameters in training and stabilizes its convergence. 
In contrast, hyper-parameter tuning and unstable environments are still big challenges for the majority of state-of-the-art DRL models such as DDPG~\cite{henderson2018deep}.
Another advantage of SAC is its robustness in handling multi-dimensional tasks. 
High-dimensional tasks are generally challenging to deal with for DRL model due to a phenomenon known as the curse of dimensionality~\cite{curseofdimension}. 
However, due to the high sample efficiency boosted by entropy maximization, SAC has demonstrated to perform very well in high-dimensional tasks with up to 21 action dimensions~\cite{sac2018}. Specifically, SAC is demonstrated to work well in the design of autonomous robots where the actions of multiple parts of the robot must be decided simultaneously. 
%On the other hand, SAC is not heavily dependent on manual hyper-parameter tuning~\cite{sac2019}, which makes it more stable to converge and achieve similar performance across various random seeds, especially in high-dimensional tasks. 
As we discuss later, we use this feature of SAC as our advantage to deal with large discrete action sets in massive MIMO user scheduling.

% A primary feature of the SAC model is the \emph{entropy regularization} concept to balance \textit{"exploration versus exploitation"}. 
% The actor component in the SAC model whose job is to maximize the expected reward, also tries to maximize the entropy. The latter forces the model to explore more widely to find more promising results. In the case that multiple promising actions are found, the policy will choose them with equal probability~\cite{sac2018}. SAC is shown to improve the learning speed over state-of-the-art DRL methods. Additionally, SAC has much less sensitivity to hyper-parameter tuning than its other counterparts. \textcolor{red}{Add more details and explain through the figure.}

\subsection{\name{} Scheduler Core Design}
\label{subsec:core_sac}
In this section, we adapt the discretized SAC algorithm~\cite{sac2019} to formulate and build a Markov Decision Process (MDP) model to solve the user scheduling problem in massive MIMO networks.

\textbf{State space.} We define the state space of user $l$ at TTI $t$ as $\textit{$s^l_t$}:= [\textit{$\gamma^l_t$},\textit{$f^l_t$},\textit{$g^l_t$}] \in \mathcal{S}:=[\Gamma,\mathcal{F},\mathcal{G}]$, where \textit{$\gamma^l_t$} indicates maximum achievable spectral efficiency of user $l$ at TTI $t$, \textit{$f^l_t$} indicates the total amount of transmitted data by user $l$ up until TTI $t$, and \textit{$g^l_t$} is the user group label of user $l$ at TTI $t$.
The value of~\textit{$\gamma^l_t$} can be calculated as the spectral efficiency of user $l$ in SU-MIMO, where only user $l$ is scheduled at TTI $t$. 
The users with the same user grouping label \textit{$g^l_t$} have low channel correlation so they are preferred to be scheduled together. 
We will introduce more details on the user grouping strategy in~\S\ref{subsec:ug}.

\textbf{Action space.} The action space set $\mathcal{A}$ consists of discrete values, encoding the user-selection decision. 
We denote the action at time $t$ as $\textit{$a_t$} \in \mathcal{A}$. 
Due to its combinatorial nature, the action set grows exponentially with the number of users in the system. 
For instance, with a total of $L$ users available, any number of users between 1 and $N_{max}$ can be scheduled at each TTI $t$, and thus the total number of possible selections is $\sum_{i=1}^{N_{max}}{L \choose i}$.% \textcolor{brown}{and $\mathcal{A}$ consists of $\sum_{i=1}^{N_{max}}{L \choose i}$ discrete values from 0 to $\sum_{i=1}^{N_{max}}{L \choose i}-1$.} 
%Because neural networks require fixed-size inputs and outputs, $N_{max}$ has to be fixed during training and inference. In massive MIMO there is a theoretical bound on $N_{max}$, and that bound is the number of antennas at the base station~\cite{bjornson2016myths}.\textcolor{cyan}{How is an action represented in your model? Is it a binary vector similar to x in Eq 7?}\textcolor{brown}{The output from SAC is a discrete value $\textit{$a_t$} \in \mathcal{A}$ and we transfer it to binary vector x in Eq 7 which represents user selection decision.}
%Then we can decide the user and modulation selection based on action value we received from model. 

\textbf{Reward.} Our ultimate objective for resource scheduling is to maximize both the system's spectral efficiency and fairness among users. 
By system spectral efficiency, we refer to the sum rate achieved by all users scheduled together at TTI $t$. 
We use a normalized version of this quantity expressed by \textbf{$\gamma^{total}_t$}. 
The normalization factor is calculated as follows. 
We measure the achievable rates for each user in the system if that user were scheduled individually (SU-MIMO). 
We then use the sum of the $N$ largest rates out of the total $L$ users as the normalization factor. This will guarantee a value in [0, 1] which then can be used in the reward function.
To quantify fairness, we use Jain’s fairness index (JFI)~\cite{jain1989corr}, which can be expressed at each TTI $t$ as

\begin{equation}
    JFI_{t} = \frac{\left ( \sum_{l=1}^{L} f_l^t \right )^{2}}{L\sum_{l=1}^{L}\left ( f_l^t \right )^{2}}.  \label{eq:jfi}
\end{equation}
%
%\textcolor{red}{Qing: I don't understand the bounds of the summation in the formula. Why is it from k to L?}
%where $f_k^t$ is latest user selection history of user $k$ at TTI $t$ and $L$ is total number of users in the cell.
%and minimum scheduled users' signal-to-noise ratio (SNR). 
%In wireless communications, SNR is a measure of the strength of the desired signal relative to background noise (undesired signal). The reason why we choose minimum scheduled user's SNR as one item in reward function is that we should ensure quality of all transmitted signals, not that the more transmitted users, the better MAC scheduler performance. 

%and \textbf{$SNR^{min}_t$} to indicate minimum SNR of selected users. 
As such, we include the normalized spectral efficiency and the JFI in the reward function of the MDP model.
The reward $\textbf{$R_t$}$ achieved at TTI $t$ can be then formulated as

\begin{equation}
R_{t} = \beta \gamma^{total}_{t}+(1-\beta) JFI_{t}.%+\gamma SNR_{t} 
\label{eq:reward}
\end{equation}

In~\eqref{eq:reward}, $\beta$ determines the relative importance of each item in the reward function based on the preference of the system operator. Note that, both items are the range $[0,1]$ so that we can effectively adjust their weights in the reward function with parameter $\beta$.
%\textcolor{red}{SS: Why two weights? Since this is relative to each other, with one weight is probably sufficient?} 
%The weights are a design choice that determine the relative importance of each item in the reward function based on the preference of the system operator. 
%By changing coefficients, we can make scheduler feature-orientated. (e.g. if $\beta$ is much bigger than $\beta$ and $\gamma$, achieving maximum system spectral efficiency is the first priority of MAC scheduler and it's spectral-efficiency-orientated)

%In MAC resource scheduling, the action space is discrete that tells scheduler which users and modulation schemes are selected. 

\subsection{Discrete Action SAC Design}
 \label{subsec:discrete}
Originally, SAC is a continuous action space model and thus, it cannot be directly applied to the massive MIMO user scheduling problem. 
There are existing discrete action space models, such as DQN~\cite{mnih2013playing} and Double DQN~\cite{van2016deep}, that could potentially be used to solve the problem. 
But as we will show in \S\ref{sec:exp}, none of these methods can handle the large action set
%\textcolor{brown}{We should use "large action set" instead of "large action space". For action set, it includes all user combinations which is large. But for action space of our model, it is just a number representing the selected user set. It's easy to make reviews confused if we use the same word here and in system model section. One review from INFOCOM is about this issue. And I found some paper discuss this problem using "action set" so I bring it here.} 
in massive MIMO user scheduling. 
Note that, the discrete action space set in multi-user scheduling in massive MIMO increases exponentially as the number of users grows. 
For example, with $M=64$ BS antennas and $L=64$ single-antenna users, or simply a $64 \times 64$ network size, and $N_{\mathrm{max}}=16$ %\textcolor{red}{You were calling the maximum number of users just $K$ before instead of $K_{max}$ ... please choose one and use it consistenlty} 
in each TTI, the action set size has up to $\sum_{i=1}^{16}{64 \choose i}\approx7\times10^{14}$ actions.

Several recent works have attempted to solve the large discrete action space problem by discretizing the continuous-control-based DRL model.
In this direction,~\cite{dulac2015drl} combines DDPG with KNN to solve problems with large discrete action sets (e.g., recommender systems and language models).
More precisely, a KNN approximation \cite{muja2014knn} is used because of its agile search in logarithmic time. 
Its fundamental idea is to first generate a so-called proto continuous action (i.e. a real number in $[-1,1]$) from the continuous action space DRL model. 
Then, KNN is used to calculate the $l^2$-norm between the proto action with actions in the discrete space represented by integer numbers corresponding to different actions, sort them in ascending order, and pick the first $\mathrm{K}$ ones. Here, $\mathrm{K}$ is a system hyper-parameter.
Finally, after comparing the Q values of these $\mathrm{K}$ discrete actions in the critic network, the one with the highest Q value is chosen as the final action. 
% Inspired~\cite{dulac2015drl}, we replace DDPG with SAC which is more superior in learning and convergence performance and utilize KNN to discretize it.
%\textcolor{red}{Are we doing exactly the same thing as the cited paper? Or we do a bit differently? We should make that clear.}\textcolor{brown}{Qing: I changed the order of statements and mentioned it after introduction of this work.} 
Similarly, we propose to augment the SAC model with a KNN approximation model that can map the continuous action space to a discrete one. However, the model in~\cite{dulac2015drl} is shown to be effective for tasks with up to one million actions, far below the number of scheduling actions encountered in a large massive MIMO network. Next, we propose an idea to scale the feasibility of the model to much larger action sets.

\subsection{Dimension Division} 
\begin{figure}[t]
      \centering
      \includegraphics[width=0.55\textwidth]{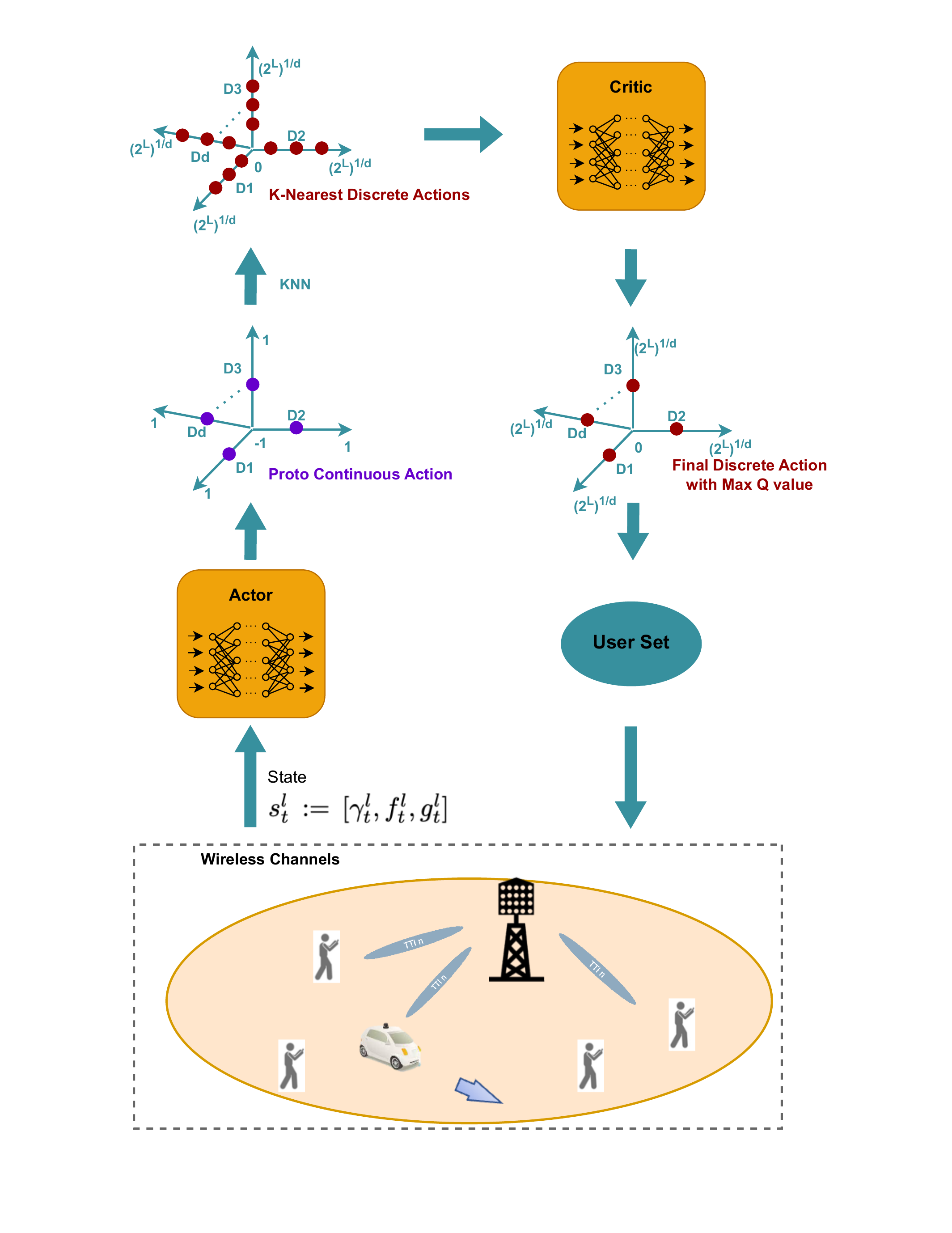}
      \caption{SMART Architecture.} 
      \label{fig:SMART_arch}
\end{figure}

One major drawback of mapping continuous actions to discrete actions is the \textit{decision accuracy loss}. The reason is that, as the size of the discrete action set increases, the corresponding distance between discrete actions in the continuous domain will become extremely small. 
The precision of each discrete action when mapped from a continuous action space in the range $[-1,1]$ is equal to ${(1-(-1))}/{2^L}$, where $2^L$ is the total number of discrete actions. 
When this precision is smaller than the network output precision, it will lead to decision accuracy loss. This precision loss prohibits scaling up the size of the discrete action set.
In order to improve the scalability of our model to much larger action sets, i.e. larger number of users, we propose a novel strategy that we call \textit{dimension division}, where we break up the action space into multiple dimensions.
As discussed in~\S\ref{sec:SAC}, high-dimensional tasks are generally challenging to deal with in DRL models. But here, we particularly rely on the strength of the SAC model in handling multiple dimensions. The difference in our approach is that we use this strength in a multi-dimensional discrete action space.
With $D$ dimensions, we can reduce the number of actions in each dimension from $2^L$ to $(2^L)^{1/D}$ actions. As such, mapping precision is also changed from $(1-(-1))/2^L$ to $(1-(-1))/(2^L)^{1/D}$ in each dimension.
Based on this strategy, the continuous-action DRL model will generate proto actions in $D$ dimensions.
We apply the approximate KNN to each proto action to generate the K nearest discrete actions in each dimension.
Finally, the critic network will pick the discrete action with the maximum Q value to form the final action (i.e. an integer number between $1$ and $2^L$).
This final discrete action is then mapped to a specific user combination from all possible combinations of $L$ users to be scheduled.
Fig.~\ref{fig:SMART_arch} illustrates the proposed workflow. 
In general, to scale up the number of supported users, it is important to strike a balance between the number of dimensions and the size of each dimension. In~\S\ref{sec:exp}, we demonstrate that the \name{} scheduler is able to perform well with a number of users as high as $L=128$ whereas DDPG is unable to converge in that scenario.
% Our proposed scheduler workflow is depicted in Fig.~\ref{fig:flow}.

% \begin{figure}[t]
%       \centering
%       \includegraphics[width=0.37\textwidth]{figures/flow.png}
%       \caption{\name{} massive MIMO Scheduler Workflow} 
%       \label{fig:flow}
% \end{figure}

\subsection{User Grouping}
\label{subsec:ug}
Previous works on DRL-based massive MIMO scheduling~\cite{guo2020globecom,chen2021eusipco} use the full channel matrix as the input to their DRL model. The size of the channel matrix is $2\times M\times L$. The factor of 2 denotes the real and imaginary components of the channel estimate since neural networks are usually designed and trained for real values. As the size of the system $(M, L)$ increases and correspondingly the input size of the DRL model grows, the model convergence becomes more difficult. In order to scale up the model to support large network sizes, the input size must be reduced. To reduce the input size, we adopt the user grouping labels calculated from the inter-user channel correlation matrix to guide the DRL model.  

%To train a DRL model for user scheduling, the model requires the knowledge of the channel. However, as the size of the system $(M, L)$ grows, the input size of the DRL model increases. 
%%Moreover, the input size also depends on the number of subcarriers in OFDM systems. 
%Additionally, considering the real and imaginary component of complex channel doubles the size of the input since neural networks are usually designed and trained for real values. Thus, the use of the raw channel matrix as the input to the DRL-based scheduler will quickly become infeasible. Previous works on DRL-based massive MIMO scheduler~\cite{guo2020globecom,chen2021eusipco} use the full channel matrix as the input to their DRL model. However, the large size of the input usually inhibit the convergence of the model. Thus, the complexity of the channel matrix must be reduced. In order to solve the channel size problem and alleviate the complexity of the state space in our DRL model, we adopt the user grouping labels calculated from inter-user channel correlation matrix to guide the DRL model. 

The inter-user channel correlation matrix measures the correlation between each pair of users in the network. Specifically, it is calculated as

\begin{equation}
     c_{i,j} = \left| \left\langle \frac{\mathbf{h_i}}{\left\| \mathbf{h_i} \right\|_{2}}, \frac{\mathbf{h_j}}{\left\| \mathbf{h_j} \right\|_{2}}\right\rangle \right|=\frac{\left| \mathbf{h_i}^{\text{H}} \mathbf{h_j} \right|}{\left\|  \mathbf{h_{i}}\right\|_{2}\left\| \mathbf{h_j} \right\|_{2}}
     \label{eq:corr}
\end{equation}
where $\mathbf{h_i}$ and $\mathbf{h_j}$ are channel vectors of user \textit{i} and user \textit{j} in channel matrix $\mathbf{H}$ and $c_{i,j}$ is their channel correlation. 
% \textcolor{red}{The algorithm can probably be written in a better for with a `while' loop instead of this nested conditionals ... but maybe not too essential to change now.}

%In order to reduce the complexity of state space, we need to use another indicator instead of channel matrix to represent channel correlation. 
To reduce the complexity of the channel matrix, we adapt a similar user grouping method with~\cite{yang2018spawc}, as shown in Algorithm~\ref{alg1}. 
The algorithm uses the inter-user channel correlation matrix calculated through equation~(\ref{eq:corr}) to partition users with low correlation into separate sets, where the partitioning threshold is $c_{th}$. 
%They use equation~\ref{eq:corr} to calculate inter-user channel correlation.
%The algorithm of user grouping is listed in Algorithm ~\ref{alg1}. 
During grouping, users in the same group (less correlated users) are assigned the same label.
As discussed in~\S\ref{subsec:core_sac}, we only then need to assign a user group label to each user in the state space instead of its complete channel vector. 
With user grouping labels as input of the DRL model, the state space size will be significantly reduced. 
As an example, in a $64\times 64$ network size, at each TTI, the state of each user includes three variables: maximum achievable spectral efficiency, the total amount of transmitted data by the user, and user group label. 
Thus, the total state space size is 192. 
However, without user grouping, the real and imaginary parts of the raw channel matrix must be fed to the DRL model separately, which leads to a state space size of 8320. 
%With maximum achievable spectral efficiency and the latest fairness value, the state space size will be 8320. 
Such large-scale inputs will lead to complicated neural network structure, high computation complexity in model updating, and excessive running time (cf. ~\S\ref{subsec:er}).

%  \begin{figure}[t]
%     \begin{minipage}{0.48\textwidth}
%     \begin{algorithm}[H]
%     \DontPrintSemicolon
%     \KwIn{Channel matrix at TTI $t$: $H_t$, user set $U$ and channel correlation threshold: $C_{th}$  
%     }
    
%     \KwOut{User group set $G$}

%   \For{User $u$ in user set $U$}
%   {
%         \If{No user group}
%         {
%             User $u$ belongs to the first user group $u_1$ and $G$ = \{$u_1$\}
%         } \else
%         { 
%             \For{User group $g_i$ in user group set $G$}
%             {
%                 Calculate user $u$ channel correlation with each user in $g_i$ using Eq~(\ref{eq:corr})
                
%                 \If{All channel correlation $< C_{th}$}
%                 {
%                     User $u$ belongs to $g_i$
%                 }
%             }
            
%             \If{User $u$ not belongs to any existing user group}
%             {
%                 User $u$ belongs to new user group $g_n$
%                 Add $g_n$ to current user group set $G$
%             }
        
%         }
        
%   }
   
%     \Return User group set $G$
 
%     \caption{User Grouping Algorithm~\cite{yang2018spawc}}
%     \label{alg1}
%     \end{algorithm}
% \end{minipage}
% \vspace{-1.5em}
% \end{figure}

 \begin{figure}[t]
    \begin{minipage}{0.48\textwidth}
    \begin{algorithm}[H]
    \caption{User Grouping Algorithm}
    % \DontPrintSemicolon
    \begin{algorithmic}[1]
    \REQUIRE{Channel matrix at TTI $t$: $H_t$, user set $\mathcal{L}$ and channel correlation threshold: $c_{th}$} 
    % }
    
    \ENSURE{User group set $G$}
    
    % \STATE Calculate channel correlations of all UE pairs $c_{i,j},\forall i,j \in \mathcal{L}$ using Eq~(\ref{eq:corr})
    % \STATE Initialize $G = \emptyset$
    % \IF {$c_{i,j} < c_{th},\forall i,j \in \mathcal{L}$}
    % \STATE $G = \mathcal{L}$
    % \ELSE
    %     % Group UE $i_0 \in N$ : $c_{i_0,j_0} < c_{th}, \forall j_0 \in N$ and $j_0 \neq i_0$ as $G_0$. Let $G^c_0 = N\setminus N_0 \neq\emptyset$
    %     \STATE Group all UEs whose correlations with other UEs in $\mathcal{L}$ are smaller than $c_{th}$ as $G_0$. Let $G^c_0 = \mathcal{L}\setminus G_0 \neq\emptyset$
    %     \WHILE{$G \neq \mathcal{L}$}
    %         \STATE Random pick UE $i \in G^c_0$ and construct user group $G_{i}$ 
            
    %         \STATE Iteratively search in $G^c_0$ to find all UEs whose channel correlations with all existing UEs in $G_{i}$ are smaller than $ c_{th}$ and add it to $G_{i}$
    %         \STATE Update $G^c_0 = G^c_0\setminus G_i$
    %         \STATE User group $G_i = G_{i}\cup G_0$ and add $G_i$ to $G$

    %     \ENDWHILE
    \STATE Calculate channel correlations of all UE pairs $c_{i,j},\forall i,j \in \mathcal{L}$ using Eq~(\ref{eq:corr})
    \STATE Initialize $G = \emptyset$
	\STATE Let $\mathcal{L}^c = \mathcal{L}$
	\WHILE{$\mathcal{L}^c \neq \emptyset$}
	    \STATE Random pick UE $i \in \mathcal{L}^c$ and add to the empty user group $G_{i}$ 
	    
	    \STATE Iteratively search in $\mathcal{L}^c$ to find all UEs whose channel correlations with all existing UEs in $G_{i}$ are smaller than $ c_{th}$ and add them to $G_{i}$
	    
	    \STATE User group $G = G\cup \{G_{i}\}$

	    \STATE Update $\mathcal{L}^c = \mathcal{L}^c\setminus G_i$
	    
	\ENDWHILE
        % \IF {$G_1\cup G_2 = N$}
        % \STATE done
        % \ELSE
        % \STATE Repeat user grouping procedure (1) - (3) until the union of $n$ user groups equals $N$
        % \ENDIF
    %\ENDIF
    \RETURN User group set $G$
    \end{algorithmic}
    \label{alg1}
    \end{algorithm}
\end{minipage}
\vspace{-1.5em}
\end{figure}

\subsection{Scheduling Across RBs}
\label{subsec:scalibility}
\begin{figure}[t]
    \centering
    \begin{subfigure}[b]{0.35\textwidth}
    \centering
      \includegraphics[width=\textwidth]{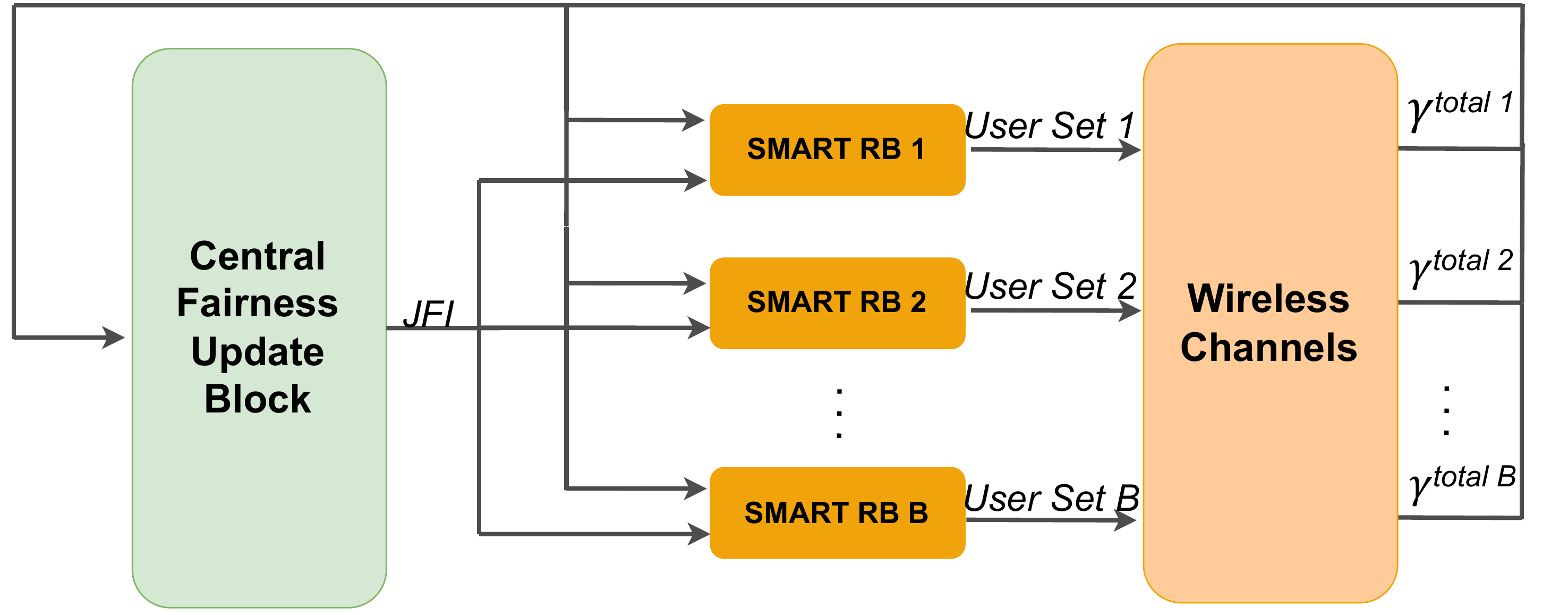}
      \caption{\name{}-MA}
      \label{fig:ma}
    \end{subfigure}
    \hfill
    \begin{subfigure}[b]{0.35\textwidth}
    \centering
      \includegraphics[width=\textwidth]{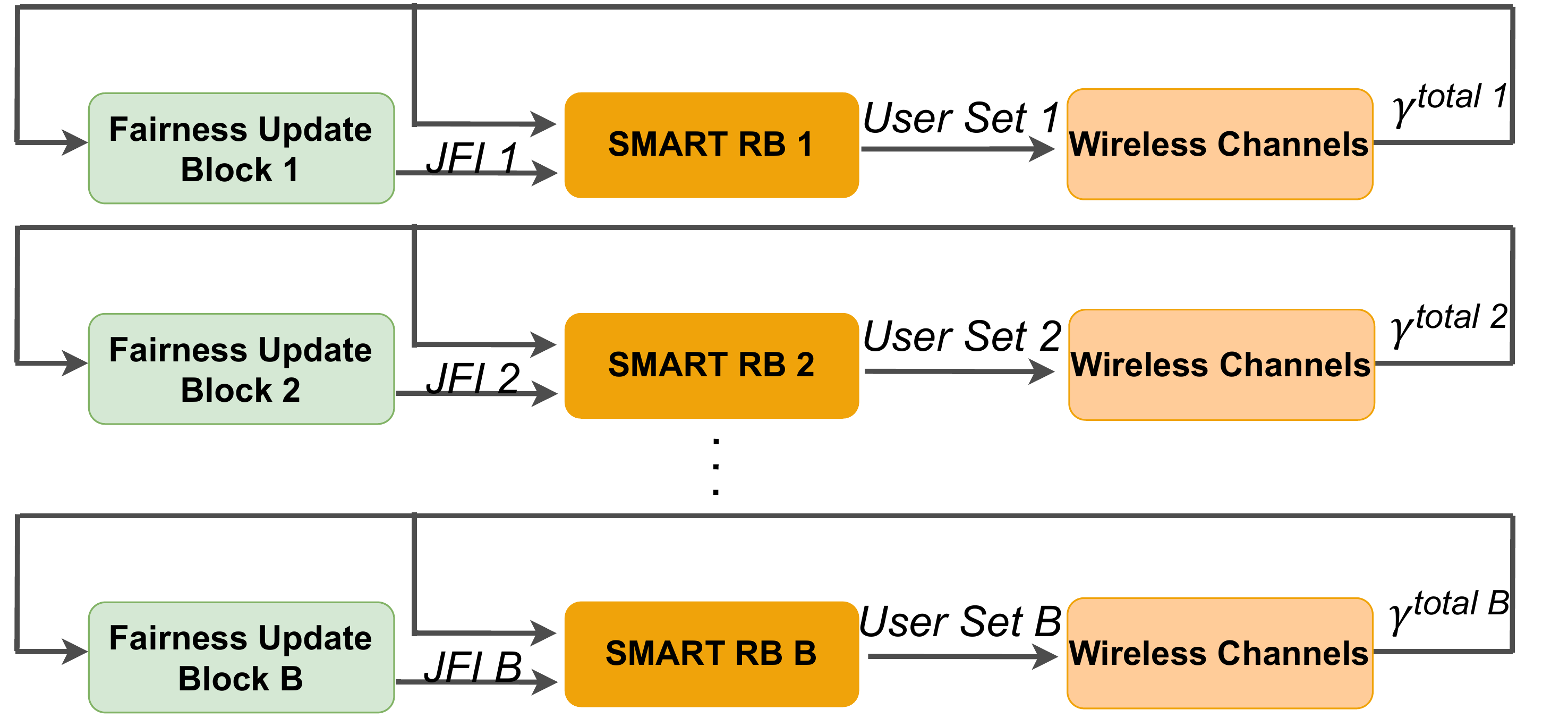}
      \caption{\name{}-SA}
      \label{fig:sa}
    \end{subfigure}
    \caption{Fully independent \name{} (a), and multi-agent \name{} frameworks (b) for scheduling users across RBs.} 
    \label{fig:sa_ma}
\end{figure}

As mentioned in \S\ref{sec:system_model}, user channel quality varies significantly across RBs. Consequently, the channel correlation among users varies across the RBs as well.
This leads to different optimal scheduling solutions for each RB. 
However, the scheduling decision on each RB will affect the decision on other RBs, particularly as it relates to rate fairness. 
Since the goal is to maximize both system spectral efficiency and fairness for the whole system, as expressed in equations (\ref{eq:max_tp_multi_rb}) and (\ref{eq:pf_multi_rb}), the optimal scheduling on all RBs needs to be jointly considered. 
One way to model this problem is to have independent SMART frameworks, as described in \S\ref{subsec:core_sac}-\S\ref{subsec:ug}, to make decisions on each RB, with the additional modification that each framework uses the decision from other frameworks running on other RBs to calculate the new fairness in its state space and the new reward, akin to the formulation of weighted rates in Eq.~\eqref{eq:pf_multi_rb}. A block diagram of such a model is depicted in Fig.~\ref{fig:ma}.
This model can be regarded as a cooperative multi-agent DRL framework where each SMART framework responsible for a different RB acts as a separate agent that shares its decisions with other agents. We refer to this overall model as \name{}-MA. %\textcolor{red}{It is formulated as~\eqref{eq:pf_multi_rb} but replacing $p_{l,b}^{t}$ with $\sum_{b}^{B}\ p_{l,b}^{t}$ in~\eqref{eq:pf_mid} as discussed in~\S\ref{sec:system_model} and we refer to it as \name{}-MA.}
In \name{}-MA, agents of RBs are jointly optimized. The instantaneous spectral efficiency of each user is aggregated from all RBs and JFI is updated based on a global user scheduling decision rather than an individual RB's decision. Consequently, the \name{} agents of all RBs share the same reward function and engage in cooperative learning.
However, multi-agent DRL models are known to be difficult to converge, especially as the number of agents scales up~\cite{4445757}. 
We demonstrate this by employing a multi-agent model in \S\ref{sec:exp}. 
For fading channel models, the inter-user channel correlation across RBs will be largely random, and when dealing with a large number of RBs, it is expected that the fairness across RBs will be smoothed out.
With this assumption, and given the limitation of the multi-agent model, we propose to use a fully independent model for each RB referred to as \name{}-SA and depicted in Fig.~\ref{fig:sa}.
In the \name{}-SA, an independent \name{} DRL model is implemented for each RB. Each RB possesses its own distinct state space (not depicted in the diagram) and generates a scheduled user set specific to that RB. Based on the user scheduling decision made by the model, selected users are allocated resources within the wireless environment, and the instantaneous spectral efficiency $\gamma^{total}$ of each scheduled user can be determined. Sequentially, the accumulated amount of transmitted data and JFI are updated in the respective fairness update block.

%\textcolor{red}{It is formulated as~\eqref{eq:pf_multi_rb} and we refer to it as \name{}-SA.} 
% The frameworks of the fully independent and multi-agent DRL models are shown in Fig.~\ref{fig:sa_ma}. We refer to them as \name{}-SA and \name{}-MA, respectively.
%\textcolor{red}{The frameworks of \name{}-MA and \name{}-SA are shown in Fig.~\ref{fig:sa_ma}. In the \name{}-SA, an independent \name{} DRL model is implemented for each RB. Each RB possesses its own distinct state space (not depicted in the diagram) and generates a scheduled user set specific to that RB. Based on the user scheduling decision made by the model, selected users are allocated resources within the wireless environment, and the instantaneous spectral efficiency $\gamma^{total}$ of each scheduled user can be determined. Sequentially, the accumulated amount of transmitted data and JFI are updated in the respective fairness update block. $\gamma^{total}$ and JFI are used to calculate the reward, which is then fed back to \name{} agent. In contrast to \name{}-SA, \name{} agents of RBs in \name{}-MA are jointly optimized. The instantaneous spectral efficiency of each user is aggregated from all RBs and JFI is updated based on a global user scheduling decision rather than an individual RB's decision. Consequently, the \name{} agents of all RBs share the same reward function and engage in cooperative learning.}

In \S\ref{sec:exp}, we demonstrate the effectiveness of \name{}-SA for a large number of RBs in getting close-to-optimal results.

%%%%%%%%% EXPERIMENTS
\section{Performance Evaluation}
\label{sec:exp}
%\textcolor{red}{Qing: Some general questions: 1) Can we use goodput instead of spectral efficiency? 2) Does channel changes across RBs in your experiments? 3) Why despite mobility spectral efficiency remains flat?}
In this section, we perform a comprehensive evaluation of our proposed scheduler design. We compare \name{} with multiple different schedulers with respect to their achieved normalized spectral efficiency and JFI in various channel conditions. We also provide a comparison of the computational complexity of our DRL-based scheduler with other methods and discuss the feasibility of our scheduler in real-time 5G settings.
%In order to be a thorough evaluation, we will simulate in two scenarios: MU-MIMO and Massive MIMO. For MU-MIMO, we consider a MIMO BS equipped with 8 antennas. It services 4 UEs with 300 m of radius. In terms of massive MIMO scenario, there are totally 64 BS antennas and 16 UEs in the same range. 

\subsection{Experimental Setup}
We perform our evaluations in both simulated channels as well as real-world channels measured with a massive MIMO hardware platform. 
For simulated wireless channels, we use the Quasi Deterministic Radio Channel Generator (QuaDRiGa)~\cite{jaeckel2014quadriga} software. Specifically, we generate the 3D Urban Micro (UMi) Line Of Sight (LOS) channel model.
%as indicated in \S\ref{sec:scheduler}. 
We consider two channel scenarios: static and mobile. For static channels, we consider two different modes: 1) four user clusters, and 2) random user placement. In the mobile scenario, the base station is positioned at the center of a circular area with a radius of 300 meters. Users within this circle move in various directions at different speeds, with an average speed of 2.8 m/s. The initial positions of the users are randomly assigned, and they will bounce back into the area upon reaching the boundary.
% we let each user move with an average speed of 2.8 m/s at different directions which ensures that we can cover as many channel conditions as possible. %To simplify analysis, we are assuming only one RB available for transmission. 
We describe the experimental setup for the real-world measured channels in \S\ref{subsec:realdataset}.
We implement the system model in~\S\ref{sec:system_model} using Python. In terms of modulation scheme, we adopt 16-QAM in our wireless channel simulator and use Error Vector Magnitude (EVM) of the received constellation to derive SNR as demonstrated in~\cite{evm}.

\begin{table}[]
\centering
\caption{Simulation and Training Parameters}

\begin{tabular}{cllcl}

\hline
\multicolumn{3}{c}{\textbf{Parameter}}       & \multicolumn{2}{c}{\textbf{Value}}     \\ \hline
\multicolumn{3}{c}{Channel Model}            & \multicolumn{2}{c}{3GPP\_3D\_UMi\_LOS} \\
\multicolumn{3}{c}{System Bandwidth}         & \multicolumn{2}{c}{20 MHz}             \\
\multicolumn{3}{c}{System Carrier Frequency} & \multicolumn{2}{c}{3.6 GHz}            \\
\multicolumn{3}{c}{TTI Duration}             & \multicolumn{2}{c}{1 ms}               \\
\multicolumn{3}{c}{Modulation}             & \multicolumn{2}{c}{16QAM}               \\
\multicolumn{3}{c}{Cell Radius}              & \multicolumn{2}{c}{300 m}              \\
\multicolumn{3}{c}{UE Speed}                 & \multicolumn{2}{c}{0 \& 2.8 m/s}            \\
\multicolumn{3}{c}{Number of BS Antennas}    & \multicolumn{2}{c}{16 \& 64}             \\
\multicolumn{3}{c}{Number of UEs}            & \multicolumn{2}{c}{16 \& 64}             \\
\multicolumn{3}{c}{Batch Size}               & \multicolumn{2}{c}{256}                \\
\multicolumn{3}{c}{Actor Learning Rate}      & \multicolumn{2}{c}{5e-4}               \\
\multicolumn{3}{c}{Critic Learning Rate}     & \multicolumn{2}{c}{5e-4}               \\
\multicolumn{3}{c}{Alpha Learning Rate}      & \multicolumn{2}{c}{3e-4}               \\
\multicolumn{3}{c}{Automatic Entropy Tuning} & \multicolumn{2}{c}{True}               \\
\multicolumn{3}{c}{Optimizer}                 & \multicolumn{2}{c}{Adam}                \\
\multicolumn{3}{c}{Episodes}                 & \multicolumn{2}{c}{800}                \\
\multicolumn{3}{c}{Iterations In Episode}    & \multicolumn{2}{c}{400}                \\
\multicolumn{3}{c}{Correlation Threshold $c_{th}$ in Algorithm~\ref{alg1}}    & \multicolumn{2}{c}{0.5}     \\
\multicolumn{3}{c}{$\beta$ in Eq.~(\ref{eq:reward})}    & \multicolumn{2}{c}{0.5}                \\ \hline
\end{tabular}
\label{tb1}
\end{table} 

% 16 cores Xeon(R) Silver 4110 GPU

We run our experiments on an NVIDIA DGX A100 server~\cite{nvidiadgx}. Both actor and critic networks implement neural nets with two hidden fully connected layers and ReLU activation functions. We use the Adam optimizer~\cite{kingma2014adam} to train our DRL model in PyTorch~\cite{paszke2017automatic}. The most relevant parameters used in our simulations are shown in Table~\ref{tb1}.

\subsection{Benchmarks}
\label{subsec:bench}
In order to do a thorough comparison, we implement various scheduler models as benchmarks including classical and heuristics-based schedulers, discrete-control-based DRL schedulers, continuous-control-based DRL schedulers, and attention-mechanism-based RL schedulers. 

\textbf{Classical Scheduler:} We consider Opt-PF, Opt-MR, an approximate PF (Approx-PF) and a heuristics-based algorithm as classical schedulers. Algorithms of Opt-PF and Opt-MR are introduced in~\S\ref{sec:system_model}. Given the exceedingly high computational complexity involved in employing optimal schedulers for large-scale networks, we devise a variation of an approximate Proportional Fairness (Approx-PF) scheduler in~\cite{approx} that offers reduced complexity from Opt-PF presented in~\S\ref{sec:system_model}. The algorithmic details of this particular implementation can be found in Algorithm~\ref{alg2}. In this approach, we first calculate a weighted-rate matrix similar to Opt-PF in~\eqref{eq:pf_multi_rb} and then select $N_{max}$ users with the highest weighted rates. Consequently, the computational complexity is reduced significantly from $\mathcal{O}$($2^{L}$) to $\mathcal{O}$($2^{N_{max}}$). However, this is still too complex in large-scale networks and thus needs to be simplified further. Unlike the approximate scheduler described in~\cite{approx}, we do not consider the individual data load of each user in our work. Instead, we implement the user grouping in Algorithm~\ref{alg1} in this user subset and select the group with the most users. User grouping strategy helps Approx-PF to avoid scheduling highly inter-correlated users, thereby improving overall system performance and releasing the heavy complexity to $\mathcal{O}$($N^2$). 

As for the heuristics-based benchmark, we use the algorithm in~\cite{yang2018spawc}. This algorithm groups users based on their channel correlation and allocates power to the users in the selected group. It then proposes to schedule the groups in a round-robin fashion. We implement a variation of the scheduler proposed in~\cite{yang2018spawc}. We assume perfect power control in our model to enable fair comparison with the modified algorithm. We refer to this benchmark algorithm as RR-UG. As we demonstrate later, this algorithm, while effective in static user scenarios, becomes ineffective in highly mobile channel scenarios where channel correlations are continuously changing. We expect a similar behavior by other heuristic methods that rely on channel correlation-based user grouping.
%\textcolor{brown}{The main limitation of~\cite{yang2018spawc} is delayed user grouping in mobility scenario. The scheduler in~\cite{yang2018spawc} only executes user grouping prior to scheduling. However, in mobility scenario, both channel condition and inter-user correlation are varying over time dramatically. The initial user groups are outdated and useless. In order to make it more competitive as our benchmark in mobility scenario, we make some adjustments. Whenever it starts scheduling, it first implements user grouping. In the upcoming few TTIs, it will schedule group by group. Once it finishes scheduling all current user groups, it will perform user grouping once more and start the subsequent scheduling run. Moreover, their algorithm also performs power allocation for users. In our implementation of their model, we assume perfect power allocation to enable fair comparison with the modified algorithm. We refer to this benchmark algorithm as RR-UG.} 

\begin{figure}[t]
    \begin{minipage}{0.48\textwidth}
    \begin{algorithm}[H]
    \caption{Approximate Proportional Fairness (Approx-PF) Algorithm}

    \begin{algorithmic}[1]
    \REQUIRE{Resource block set $\mathcal{B}$, Channel matrix of resource block $b$ at TTI $t$: $H_{t,b}$ and user set $\mathcal{L}$} 
    
    \ENSURE{Scheduled user set on resource block $b$:\ $\mathcal{U}_b$}
    \STATE Calculate weighted rate $w_{l,b}^{t}$ for all $L$ users on resource block $b$ at TTI $t$ using~\eqref{eq:pf_multi_rb}
    \STATE Sort and select $N$ users with the highest weighted rate on resource block $b$ to construct a subset of user $\mathcal{N}_b$
    \STATE Do user grouping in user subset $\mathcal{N}_b$ as Algorithm~\ref{alg1}
    \STATE Find the user group $\mathcal{U}_b$ with the most users as the scheduled user set on resource block $b$ at TTI $t$
    \RETURN $\mathcal{U}_b$
    \end{algorithmic}
    \label{alg2}
    \end{algorithm}
\end{minipage}
\vspace{-1.5em}
\end{figure}

\textbf{Discrete-control-based DRL Scheduler:} There are several DRL models for discrete action spaces in the literature. We select DQN ~\cite{mnih2013playing} and Double DQN~\cite{van2016deep} with Prioritized Experience Replay Buffer (PERB)~\cite{per} as two representative discrete-control-based DRL algorithms. The study in~\cite{acer} shows a comparison of these two model with other discrete DRL models such as ACER and A3C and shows the superior performance and convergence of our selected benchmarks.
We implement both discrete-control-based DRL models as benchmarks and refer to them as PRTY-DQN and PRTY-DDQN.
To balance exploration and exploitation, we adopt the epsilon-greedy algorithm in both models. For fair comparison against other benchmarks, we tune the hyper-parameters so as to achieve the best possible performance~\cite{sac2018,ppo,acer}. Because of the simple neural network structure of PRTY-DQN and PRTY-DDQN, we adopt grid search to comprehensively identify the optimal hyper-parameters.
For PRTY-DQN, we implement 2-hidden-layer neural networks with 32 neurons in each layer. 
We use the same settings in the main network and the target network of PRTY-DDQN. 
For both models, we set the same state space, action space, and reward function as our proposed scheduler.
%to randomly improve the current knowledge of the model about each action or choose the greedy action with the most reward. 
%\textcolor{brown}{DQN's network structure adopts 2 hidden layers, adam optimizer and mean squared error (MSE) as loss function. For Double DQN, both main network and target network have the same composition as DQN. Moreover, in terms of MDP model, DQN and Double DQN retain the same state space, action space and reward function as our scheduler.} 
%\textcolor{red}{Here explain how you adapt DQN and Double DQN for the scheduling problem. For example, what reward function you use etc.}

\textbf{Continuous-control-based DRL Scheduler:} 
Similar to SAC, DDPG is also a continuous-control-based DRL model that has been used to solve optimization problems with large action sets, e.g., on massive MIMO user scheduling~\cite{guo2020globecom, dulac2015drl}. 
%However, as mentioned in~\ref{subsec:discrete}, it has server limitations on scalability to large massive MIMO network. 
To compare SAC with a DDPG-based scheduler, we replace the SAC module in our design with DDPG and use it as our benchmark. For fairness of comparison, this benchmark adopts the same dimension division strategy as our design to generate multi-dimensional scheduling actions, particularly in evaluating $64\times 64$ network size. Furthermore, we use the same state space and reward function as well as the epsilon-greedy algorithm for this benchmark algorithm as in our proposed scheduler.

\textbf{Attention-mechanism-based RL Schedulers:} 
We implement a pointer-network-based scheduler (PN) as proposed in~\cite{chen2021eusipco} in an actor-critic architecture. 
%~\textcolor{red}{Why this algorithm? Explain why this is good?} \textcolor{brown}{Qing: this paper uses PN to design user scheduler, so we implement the same one as it. For DRL model-based PN, it's popular to use actor-critic to train model.} 
The PN is used as the actor network, which consists of a long short-term memory (LSTM)-based encoder and decoder. 
The critic network is a multi-layer perceptron (MLP) and is trained using stochastic gradient descent. A limitation of this model is that the number of scheduled users needs to be fixed. Thus, in our evaluation of the PN scheduler, we set the number of scheduler users $N$ to be so that $M/N\approx 4.5$ which is shown to be the near-optimal number for the ZF beamformer~\cite{bjornson2016myths}. 
%\textcolor{red}{Perhaps a fair comparison is to also fix the number of users for SMART to compare against pointer network. For e.g. We have SMART-FIXED-UE. But let's discuss.} \textcolor{brown}{Qing: Flexible scheduled user number is our advantage. Many related work, such as~\cite{chen2021eusipco, guo2020globecom}, require a pre-defined user number, which makes them impossible to determine the number of users according to the channel quality. It will result from performance degradation.}

\textbf{Our Proposed Scheduler:}
%Our scheduler combines the SAC model with the KNN algorithm to infer scheduling decisions as described in~\S\ref{subsec:core_sac} and~\S\ref{subsec:discrete}. 
We implement two variants for our scheduler: 1) a variant with raw channel matrix as input that we call \name{}-Vanilla, and 2) a variant with user grouping labels as input (as described in~\S\ref{subsec:ug}) that we simply call \name{}. % and 3) \name{} with user grouping and modulation selection (\name{}-UM). 

In our evaluations, the Opt-PF scheduler serves as the optimal benchmark for fairness while the Opt-MR scheduler is optimal for spectral efficiency. 
For thoroughness, we first rule out the discrete DRL-based scheduler, i.e. DQN and Double DQN, due to their inability to scale to large network sizes. Second, we compare the remaining benchmarks in a medium $16\times 16$ network size and in different channel conditions. This allows comparison of the AI-based benchmarks with Opt-PF and Opt-MR schedulers when they are still in a computationally feasible range.
Lastly, we increase the size of the network to $64\times 64$, which we consider a real-world network size. 
In this network size, both Opt-PF and Opt-MR schedulers become computationally infeasible and thus, we only compare our proposed schedulers with PN, DDPG, and RR-UG.
%The modulation is fixed at QAM-16 for all schedulers we implement.

% \subsection{Real-World Dataset}
% \label{subsec:rwdata}
% \textcolor{red}{To show how to collect real-world dataset.(what scenario cases are covered, topology)}

\subsection{Results}
\label{subsec:er}
\subsubsection{Model Training and Convergence}
We trained the \name{} model, in a $64\times 64$ network size, for 800 epochs with 400 iterations in each epoch. To ensure model convergence and learning performance, we divide $8$ dimensions in action space and $256$ actions in the action set of each dimension, as discussed in~\ref{subsec:discrete}. The training takes about five hundred epochs which is when the DRL model converges. During the training process, we employ the epsilon-greedy algorithm to effectively manage the trade-off between exploration and exploitation. This is achieved by selecting random actions or utilizing learned actions that yield the highest reward. The value of epsilon denotes the probability of selecting random actions for exploration purposes. Initially, we set epsilon to 1, and gradually decrease it to zero over a span of five hundred epochs.

We also trained \name{} for a $128\times 128$ network. To deal with this extremely large action set, we break it down into 16 dimensions with 256 actions in each dimension for sufficient decision accuracy. With these parameters, we find that our DRL model can still converge. Conversely, all other RL-based benchmarks, except PN, fail to converge in this scenario. However, as we show later, the training and inference time for PN is significantly larger and its performance in terms of fairness is inferior to our scheduler. It is important to highlight that \name{}-Vanilla cannot converge in networks of this size either due to the excessive state space. This observation further emphasizes the motivation behind incorporating user grouping in \name{}.

\textbf{Convergence of PRTY-DQN and PRTY-DDQN:}
Discrete-based DRL is intuitively a suitable choice to deal with discrete combinatorial optimization problems, such as resource scheduling, by modeling them as MDPs. However, in problems with large action sets, the discrete-based DRL model is shown unable to converge during the training process~\cite{chen2021eusipco, vandewiele2020qlearning}, an effect known as the action dimension disaster~\cite{chen2021eusipco}. We also demonstrate this effect by training PRTY-DQN and PRTY-DDQN on multiple network sizes. Our experiments show that the largest network size that these models could converge is $4\times 4$, and $N_{\mathrm{max}}=2$. In this configuration, the size of the action set is 10.
%, which is small enough to make DQN and Double DQN converge like shown in Fig.~\ref{fig:lr_dqn_ddqn}. It is evident that Double DQN has faster convergence than DQN. The reason is that Double DQN adopts two separate networks to perform action selection and evaluation to overcome overestimation which makes DQN waste time exploring inefficient actions. However, the considered network size is too small for real world comparison. Nevertheless, we provide a performance comparison of DQN and Double DQN with our proposed scheduler and other benchmarks in the next section. \textcolor{brown}{Qing: Do the last two sentences have logical issue? Even though ..., ...? }
%\textcolor{cyan}{What happened to Priority DQN? The figures are still showing that.} \textcolor{blue}{Fixed}
% \begin{figure}[t]
%     \centering
%     \begin{subfigure}[b]{0.24\textwidth}
%     \centering
%       \includegraphics[width=\textwidth]{figures/lr_dqn.png}
%       \caption{}
%       \label{fig:lr_dqn}
%     \end{subfigure}
%     \hfill
%     \begin{subfigure}[b]{0.24\textwidth}
%     \centering
%     \includegraphics[width=\textwidth]{figures/lr_ddqn.png}
%       \caption{}
%       \label{fig:lr_ddqn}
%     \end{subfigure}
%     \caption{Training curves of (a) DQN and (b) Double DQN in 4$\times$4 massive MIMO network.} 
%     \label{fig:lr_dqn_ddqn}
% \end{figure}

\subsubsection{Performance Comparison in Various Network Sizes}
\label{sec:peformance}

In the testing phase, we run our simulation environment for additional 400 TTIs in the same cell and use the trained model to schedule users while recording the spectral efficiency and the JFI values across TTIs. For a fair comparison, we use the exact same channels generated as input to all benchmarks. It is important to note that partial or outdated channel information could impair the performance of the resource scheduler, particularly in scenarios involving high-speed mobility. This impacts any system that relies on the channel information for scheduling decisions and thus is beyond the scope of our work. Nevertheless, in this case, complementary methods that perform channel prediction based on the partial or outdated channel information such as the ones proposed in~\cite{channelprediction,8949454,8693948} can be used to enhance the performance of the scheduler. %\textcolor{brown}{As depicted in~\S\ref{sec:system_model}, our system model is built on the basis of full knowledge of the channel condition of all users associated with the BS. Even in practical wireless communication systems, partial observability of the channel condition is more common. It is also important to note that partial observability impairs system performance due to outdated channel information, particularly in scenarios involving high-speed mobility. Nevertheless, in this case, we can use complementary methods that perform channel prediction based on the partial channel information such as the ones proposed in~\cite{channelprediction,8949454,8693948}.}
In the following, we provide evaluation results of various benchmarks in multiple network sizes. In each network size, we plot the average spectral efficiency and JFI over all TTIs. We also display error bars in each plot indicating the minimum and maximum values of results across TTIs.

\textbf{Small network size:} We consider the $4\times 4$ network configuration in a mobile scenario, to compare the performance of PRTY-DQN and PRTY-DDQN with our proposed scheduler.
%Moreover, the reward achieved by Double DQN is more than DQN with the same reward function. 
%We also implement them in a larger network size with 8 BS antennas and 8 UEs, where total action set size is 162 and both models fail to converge. 

\begin{figure}[t]
    \centering
    \begin{subfigure}[b]{0.24\textwidth}
    \centering
      \includegraphics[width=\textwidth]{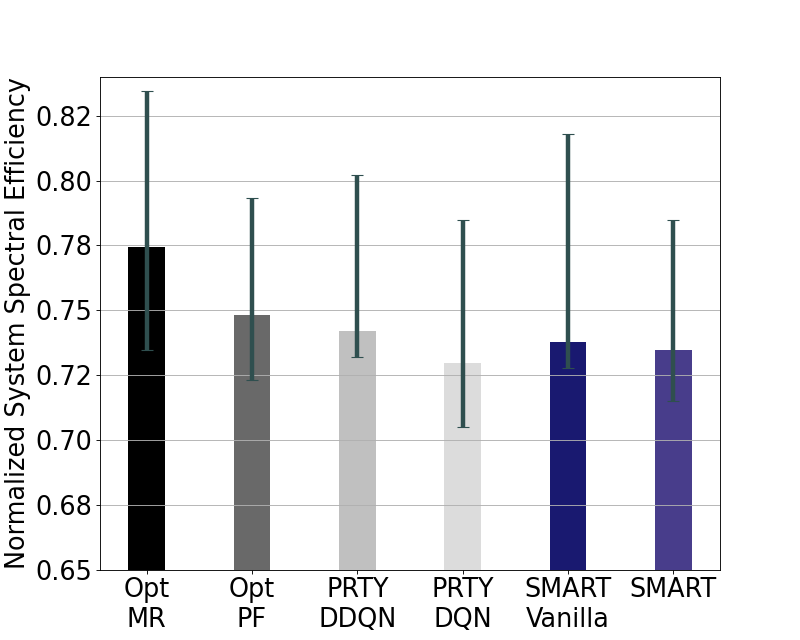}
      \caption{}
      \label{fig:se_4_mob}
    \end{subfigure}
    \hfill
    \begin{subfigure}[b]{0.24\textwidth}
    \centering
    \includegraphics[width=\textwidth]{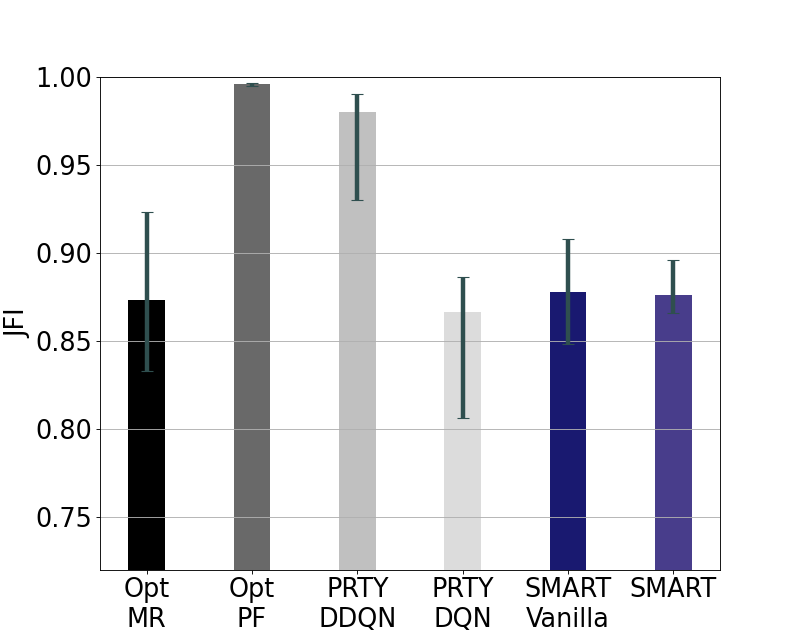}
      \caption{}
      \label{fig:jfi_4_mob}
    \end{subfigure}
    \caption{Spectral Efficiency and JFI comparison of \name{} with DQN and Double DQN in user mobility scenario and $4\times 4$ network size.} 
    \label{fig:4_2_4}
\end{figure}

Fig.~\ref{fig:4_2_4} shows that PRTY-DDQN outperforms PRTY-DQN and SMART-Vanilla on both spectral efficiency and JFI. This is due to decision accuracy loss imposed by mapping the SAC output from continuous space to discrete space in our scheduler, as discussed in~\S\ref{subsec:discrete}. However, the limitation on the scalability of PRTY-DDQN makes it impractical to use in real-world network sizes. mportantly, we observe that the performance of \name{} is almost the same as \name{}-Vanilla. This is an important finding since it shows using user grouping labels as input to our model instead of the raw channel matrix as in~\name{}-Vanilla simplifies neural network structure while not impairing model performance. %\textcolor{brown}{Applicable to all figures, the error bar displays the outliers (i.e. both extreme values) of normalized system spectral efficiency and JFI.}

% \begin{figure}[t]
%     \centering
%     \begin{subfigure}[b]{0.24\textwidth}
%     \centering
%       \includegraphics[width=\textwidth]{figures/Nor_4_SE_mob.png}
%       \caption{}
%       \label{fig:se_4_mob}
%     \end{subfigure}
%     \hfill
%     \begin{subfigure}[b]{0.24\textwidth}
%     \centering
%     \includegraphics[width=\textwidth]{figures/JFI_4_mob.png}
%       \caption{}
%       \label{fig:jfi_4_mob}
%     \end{subfigure}
%     \caption{Spectral Efficiency and JFI comparison of \name{} with DQN and Double DQN in user mobility scenario and $4\times 4$ network size.} 
%     \label{fig:4_2_4}
% \end{figure}

% \begin{figure}[t]
%     \centering
%     \begin{subfigure}[b]{0.24\textwidth}
%     \centering
%       \includegraphics[width=\textwidth]{figures/Nor_4_SE_box.png}
%       \caption{}
%       \label{fig:se_4_mob}
%     \end{subfigure}
%     \hfill
%     \begin{subfigure}[b]{0.24\textwidth}
%     \centering
%     \includegraphics[width=\textwidth]{figures/JFI_4_box.png}
%       \caption{}
%       \label{fig:jfi_4_mob}
%     \end{subfigure}
%     \caption{Spectral Efficiency and JFI comparison of \name{} with DQN and Double DQN in user mobility scenario and $4\times 4$ network size.} 
%     \label{fig:4_2_4}
% \end{figure}

\begin{figure}[t]
    \centering
    \begin{subfigure}[b]{0.24\textwidth}
    \centering
      \includegraphics[width=\textwidth]{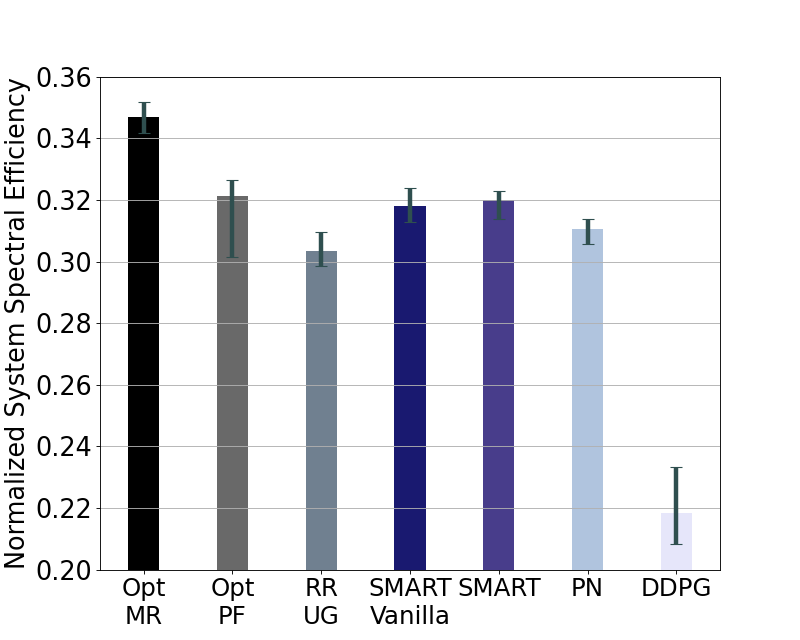}
      \caption{}
      \label{fig:se_16_4cluster}
    \end{subfigure}
    \hfill
    \begin{subfigure}[b]{0.24\textwidth}
    \centering
      \includegraphics[width=\textwidth]{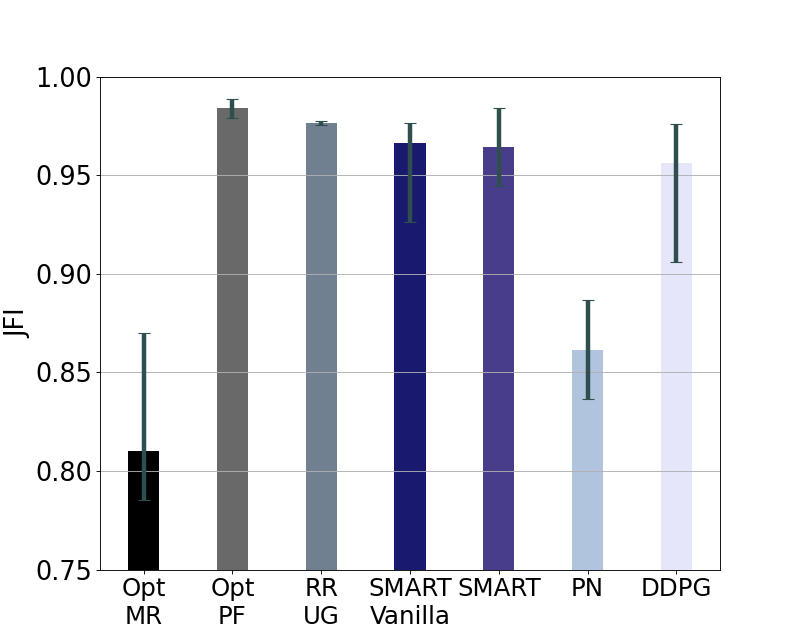}
      \caption{}
      \label{fig:jfi_16_4cluster}
    \end{subfigure}
    \caption{Spectral Efficiency and JFI comparison of \name{} and existing methods in $16\times 16$ network size and $N_{\mathrm{max}} = 4$ in 4-clusters topology.} 
    \label{fig:se_jfi_16_4clu}
\end{figure}

\textbf{Medium network size}: 
For thorough comparison of all the other benchmarks, we consider the case for medium $16\times 16$ network size, and $N_{\mathrm{max}}=4$. We only compare the benchmarks with \name{}-Vanilla for a fair comparison with other AI-based schedulers which use the raw channel matrix as input.
To be able to reason about the performance of each scheduler, we start with a toy network scenario where the users are static and placed in four clusters (4-cluster).
The users in each cluster share the same scatters and experience similar small-scale fading, and thus their channel vectors are highly correlated. Fig.~\ref{fig:se_jfi_16_4clu} shows the spectral efficiency and JFI results in the four-cluster channel mode.
It is evident from Fig.~\ref{fig:se_16_4cluster} that \name{}-Vanilla performs very close to Opt-PF scheduler, which shows \name{}-Vanilla is able to converge to the Opt-PF solution almost perfectly. 
In terms of JFI, Fig.~\ref{fig:jfi_16_4cluster} shows that \name{}-Vanilla closely follows the Opt-PF scheduler as well. 
Both schedulers underperform the Opt-MR scheduler in terms of spectral efficiency, but the Opt-MR scheduler is not doing well with respect to JFI as expected, since it is only optimizing the spectral efficiency.
Interestingly, Fig.~\ref{fig:se_16_4cluster} also shows the DDPG-based scheduler significantly under-perform \name{}-Vanilla. 
That shows DDPG fails to explore widely enough because of its sample inefficiency and therefore gets stuck in a local optimal. 
%But \name{} mitigates due to the maximization of entropy in the policy function of the SAC. 
Lastly, we observe that RR-UG achieves a good spectral efficiency and is almost close to \name{}-Vanilla. 
This is expected as the user grouping algorithm groups the users into exactly four groups based on four clusters. 
Since the users do not move, RR-UG will continue to serve each group at a time. 
The results also show that \name{}-Vanilla can learn the inter-user correlation well, despite using the raw channel matrix from each user.
PN is able to achieve near-optimal spectral efficiency but undesirable JFI. 
The reason is that PN can not deal with varying state representations of the input~\cite{nazari2018deep}. 
Specifically, sequentially selecting the users will affect the fairness in the state space of the MDP model. 
Therefore, PN fails to optimize the JFI, while still performing well in terms of achieved spectral efficiency. %However, as we show below, this will change as we move to other network topologies. %\textcolor{red}{We should mention somewhere why PN could be attractive due to it scalability.} \textcolor{brown}{I include it in benchmark PN part.}

\begin{figure}[t]
    \centering
    \begin{subfigure}[b]{0.24\textwidth}
    \centering
      \includegraphics[width=\textwidth]{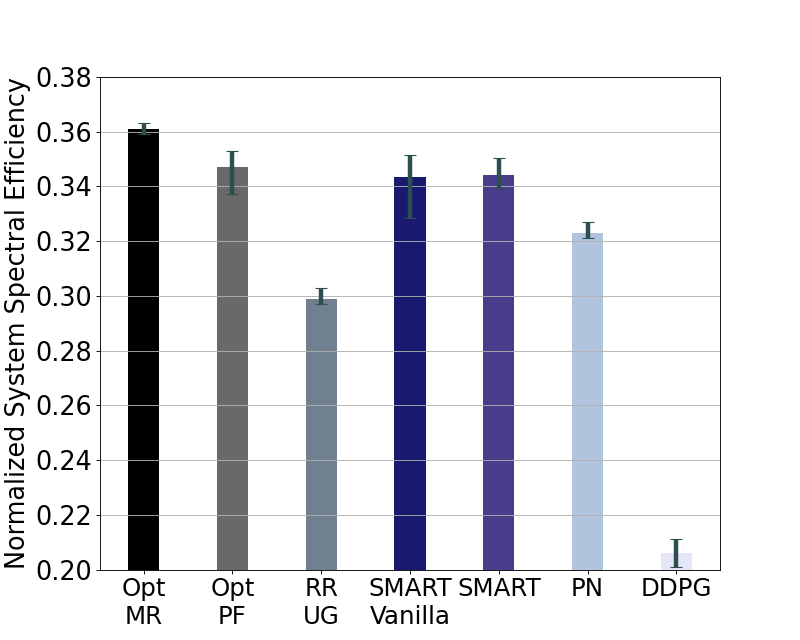}
      \caption{}
      \label{fig:se_16_random}
    \end{subfigure}
    \hfill
    \begin{subfigure}[b]{0.24\textwidth}
    \centering
      \includegraphics[width=\textwidth]{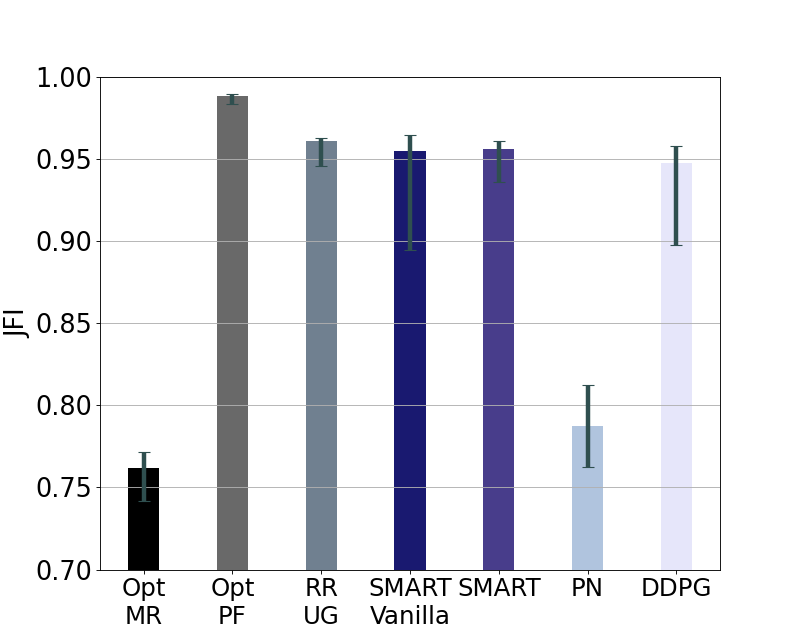}
      \caption{}
      \label{fig:jfi_16_random}
    \end{subfigure}
        \begin{subfigure}[b]{0.24\textwidth}
    \centering
        \includegraphics[width=\textwidth]{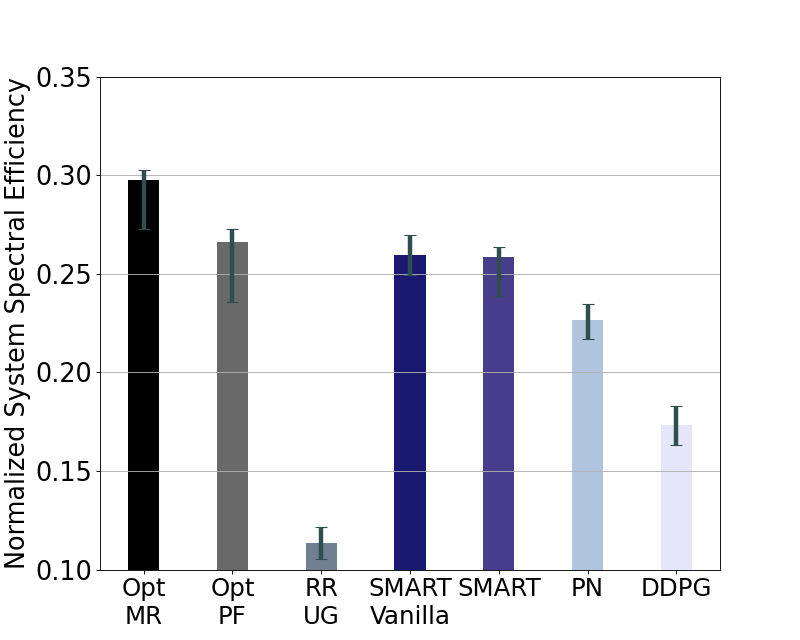}
        \caption{}
        \label{fig:se_16_mob}
    \end{subfigure}
    \hfill
    \begin{subfigure}[b]{0.24\textwidth}
    \centering
        \includegraphics[width=\textwidth]{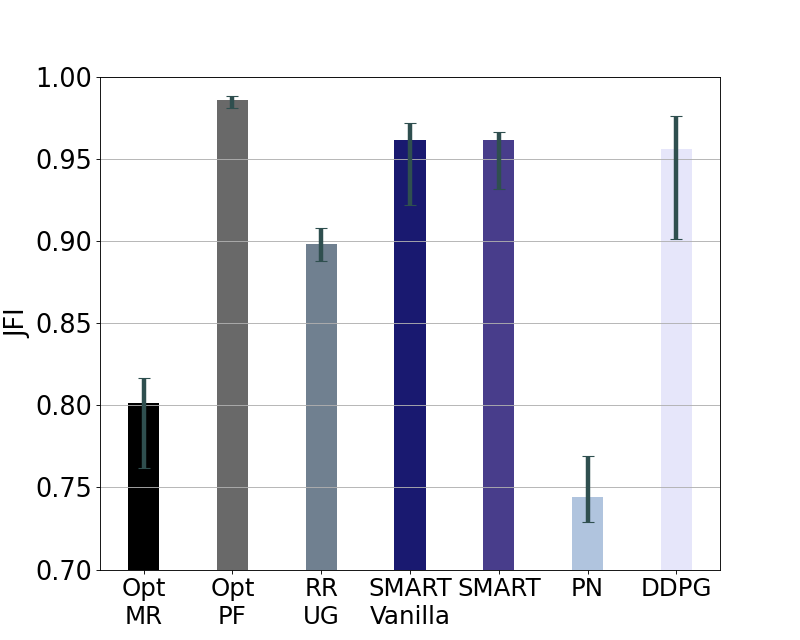}
        \caption{}
        \label{fig:jfi_16_mob}
    \end{subfigure}
    \caption{Spectral Efficiency and JFI comparison of \name{} and existing methods in $16\times 16$ network size and $N_{\mathrm{max}} = 4$ in static random user topology (a) and (b), and user mobility scenario (c) and (d).} 
    \label{fig:se_jfi_16}
\end{figure}

Figs.~\ref{fig:se_16_random} and~\ref{fig:se_16_mob} show the normalized spectral efficiency for random placement of static users in the cell and mobile users moving in random directions within the cell, respectively. 
In both scenarios, we observe that \name{}-Vanilla still performs very closely to the Opt-PF scheduler, while the DDPG scheduler significantly underperforms \name{}-Vanilla. The PN performance also slightly drops compared to the 4-cluster scenario. 
This can be attributed to the limitation of this scheduler with respect to its predefined number of selected users. 
Note that in the 4-cluster scenario, the predefined number of scheduled users for PN is exactly the same as the number of users in each user group where users have very low correlation. 
However, in the random placement scenario, this condition does not necessarily hold and the number of scheduled users by PN could be smaller or larger than the optimal set of users. The PN performance gets worse in the mobility scenario since user grouping is changing over time. For instance, PN could select user sets with high correlation in most cases. 
%\textcolor{red}{Looking at the plots, I don't see a much worse performance for PN in random and mobile scenarios!} \textcolor{brown}{Qing: not too much worse, normalized performance gap between SMART and PN from (random placement) 0.02 to (mobility) 0.05}.
%A fixed scheduled user number is difficult to adapt to varying channel condition. 

RR-UG achieves a relatively good performance in random placement topology, but it does not achieve the same level of performance as in the 4-cluster channel mode. 
The reason is that in the setups with random user locations, the user groups could include a larger number of users than $N_{\mathrm{max}}=4$, and thus the groups have to be broken into smaller subgroups to be scheduled sequentially. 
This impairs the performance of RR-UG. 
In the mobility scenario, the performance of RR-UG drops even more. 
This is due to the variations in channels and user groupings caused by mobility in each TTI. 
It shows that while RR-UG might be a favorable scheduler in static scenarios (due to its lower computational complexity as we show later), in the mobility scenarios, it does not perform that well. 
In Figs.~\ref{fig:jfi_16_random} and~\ref{fig:jfi_16_mob}, we see \name{}-Vanilla and DDPG achieve high fairness values. 
A good fairness result for DDPG is expected as fairness is accounted for in the reward function. 
Opt-MR and RR-UG do not achieve high fairness in both scenarios. 
For RR-UG, the fairness drops since the user groupings change continuously, and thus the rate fairness cannot be met efficiently despite the time fairness due to the Round-Robin scheduling of groups. It is evident that PN performs very poorly with respect to JFI, as discussed earlier. 

%\textbf{JFI Inter-user Fairness}. Fig.~\ref{fig:jfi_16} depicts JFI comparison of different TTIs under our proposed algorithm and benchmarks in 8 x 4 MU-MIMO. We find that \name{} schedulers is able to achieve comparable JFI after 400 TTIs with traditional RR and PF schedulers. The same conclusion is also be derived from fig.~\ref{fig:big_jfi}, which is in 64 x 16 massive MIMO. We don't show result of PF scheduler due to the same reason as system spectral efficiency comparison. 
 
\begin{figure}[t]
    \centering
    \begin{subfigure}[b]{0.24\textwidth}
    \centering
      \includegraphics[width=\textwidth]{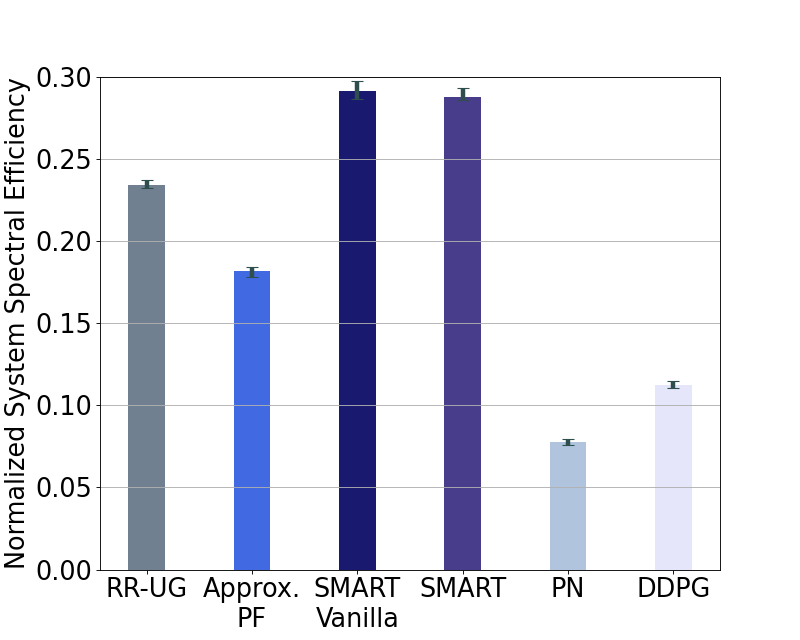}
      \caption{}
      \label{fig:se_64_random}
    \end{subfigure}
    \hfill
    \begin{subfigure}[b]{0.24\textwidth}
    \centering
    \includegraphics[width=\textwidth]{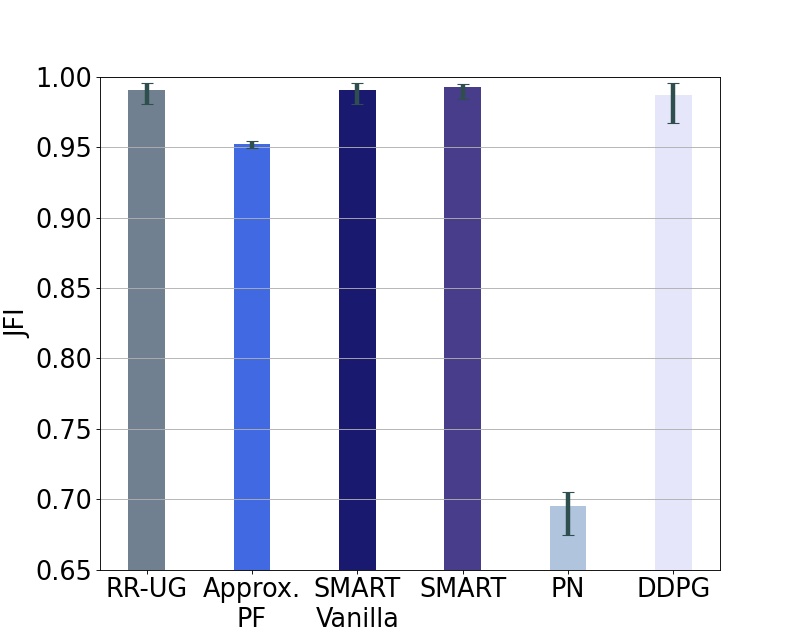}
      \caption{}
      \label{fig:jfi_64_random}
    \end{subfigure}
    \hfill
    \begin{subfigure}[b]{0.24\textwidth}
    \centering
      \includegraphics[width=\textwidth]{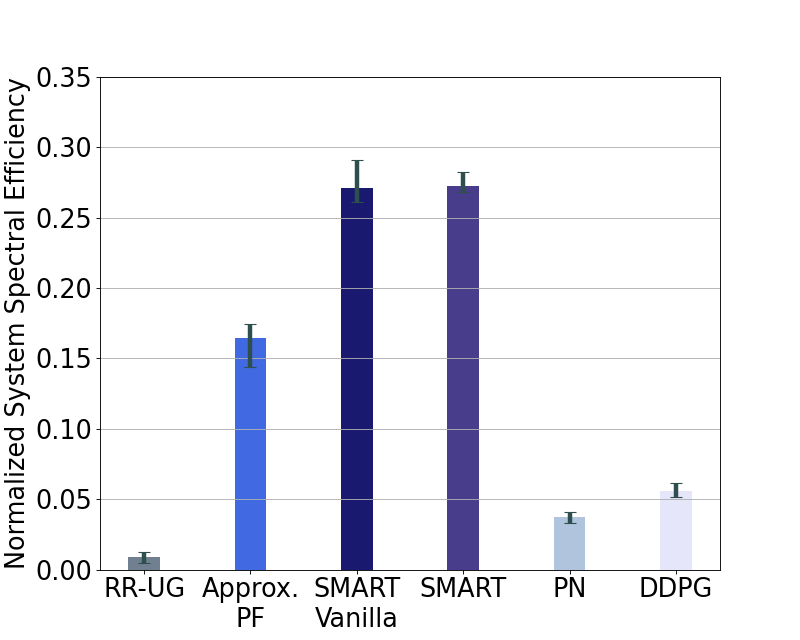}
      \caption{}
      \label{fig:se_64_mob}
    \end{subfigure}
    \hfill
    \begin{subfigure}[b]{0.24\textwidth}
    \centering
    \includegraphics[width=\textwidth]{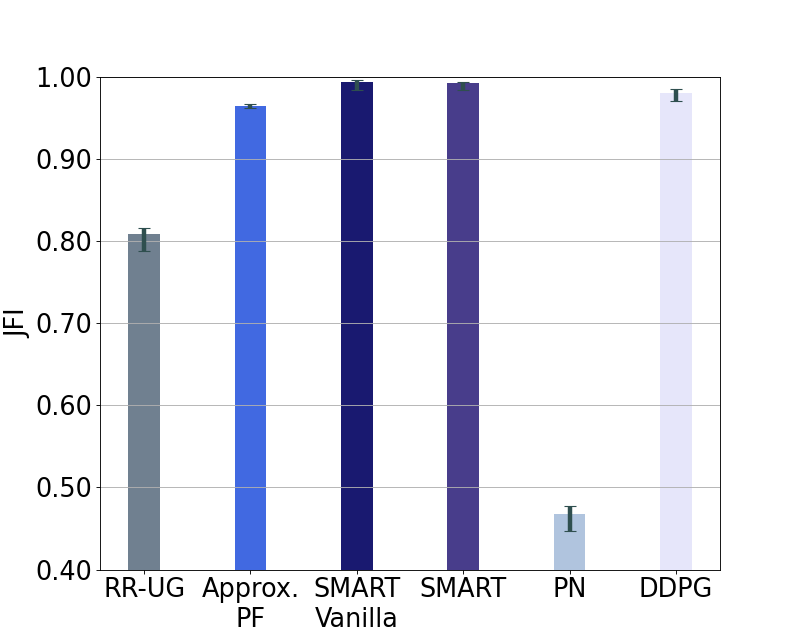}
      \caption{}
      \label{fig:jfi_64_mob}
    \end{subfigure}
    \caption{Spectral efficiency and JFI comparison of \name{} and existing methods in $64\times 64$ network size in random user topology (a) and (b), and user mobility scenario (c) and (d).}
    \label{fig:se_64}
\end{figure}

\textbf{Real-world network size}: We consider a more realistic network size with a 64-antenna massive MIMO base station\footnote{Most commercial deployments of massive MIMO include 64-antenna base stations} at the center of the cell. 
We also consider $L=64$ connected users which is also a realistic number in small cells~\cite{chen2020twc}. In this case, we assume $N_{\mathrm{max}}=64$ which means the scheduler can choose to beamform to up to all 64 users in one TTI. 
In this network size, the complexity of calculating the results for Opt-MR and Opt-PF is too high.%\textcolor{red}{as $\mathcal{O}(2^L)$}. 
Thus, we include Approx-PF as a benchmark instead of Opt-PF along with the results for \name{}-Vanilla, \name{}, PN, DDPG, RR-UG. As shown in Figs.~\ref{fig:se_64_random} and~\ref{fig:se_64_mob}, \name{}-Vanilla outperforms PN, DDPG, RR-UG, and Approx-PF. By foregoing the exhaustive search, Approx-PF aims to reduce computational complexity. However, we can see that its performance falls short compared to \name{}.
Similar to our earlier results for medium network size, the performance of RR-UG is close to \name{}-Vanilla in static random user placement but drops significantly in the mobility scenario. 
%\textcolor{brown}{Compared to small-size network size, performance discrepancy between \name{} and DDPG in real-size network is more obvious. In large-size network, we adopt dimension division as discussed in~\ref{subsec:discrete} to ensure mapping accuracy in cost of high-dimensional actions. For high-dimensional tasks, SAC's sample efficiency advantage over DDPG is more significant.} 
To enable DDPG to converge in this scenario, we applied the dimension division presented in~\ref{subsec:discrete} to its implementation. However, DDPG is unable to perform well in multi-dimensional action sets as discussed earlier. This explains the observation that DDPG does not perform well in terms of spectral efficiency.
As we observed in the small and medium networks, the performance of \name{} is comparable to that of \name{}-Vanilla in both channel scenarios. It demonstrates the effectiveness of using user grouping labels in the state space of \name{}.

% Also \name{}-UM which includes modulation selection achieves significantly better results similar to the earlier results. 
All schedulers, except PN, achieve high fairness in the static random user placement scenario. In the mobility scenario, the fairness for RR-UG also drops significantly due to varying user groupings across TTIs. Here, PN has the worst JFI for the same reason as we mentioned for the medium network size.

\subsubsection{Multi-RB Scheduling Performance}
\label{sec:scale}
%\textcolor{red}{Lots of incoherent text here. I will rewrite after new results are added.}
%When we extend scheduling from a single RB to a large number of RBs, the action taken on each individual RB is affected by others' actions only because of coordination on inter-user fairness among multiple RBs.  Spectral efficiency optimization are independent from each other. Therefore, JFI is the most important metric to evaluate performance of multiple RBs.

\begin{figure}[t]
    \centering
    \begin{subfigure}[b]{0.235\textwidth}
    \centering
      \includegraphics[width=\textwidth]{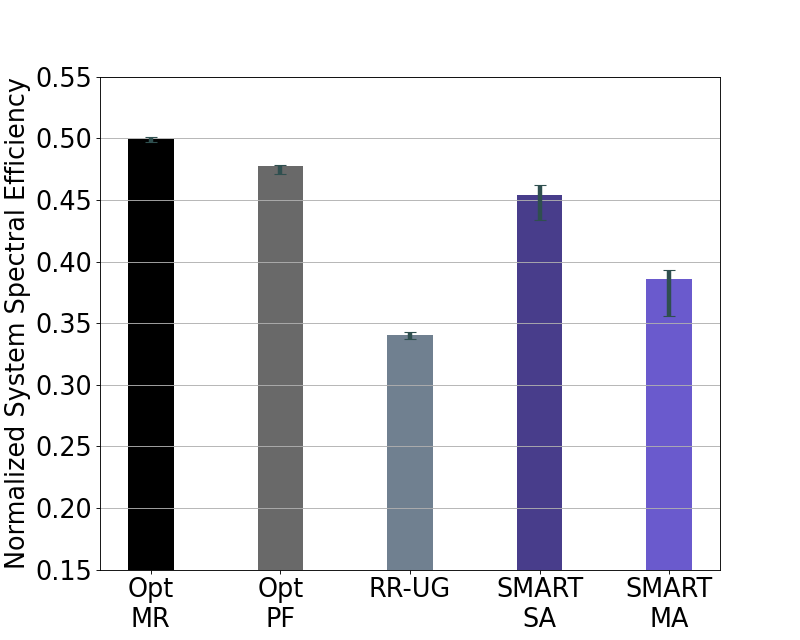}
      \caption{}
      \label{fig:2rb_se_8_random}
    \end{subfigure}
    \hfill
    \begin{subfigure}[b]{0.24\textwidth}
    \centering
      \includegraphics[width=\textwidth]{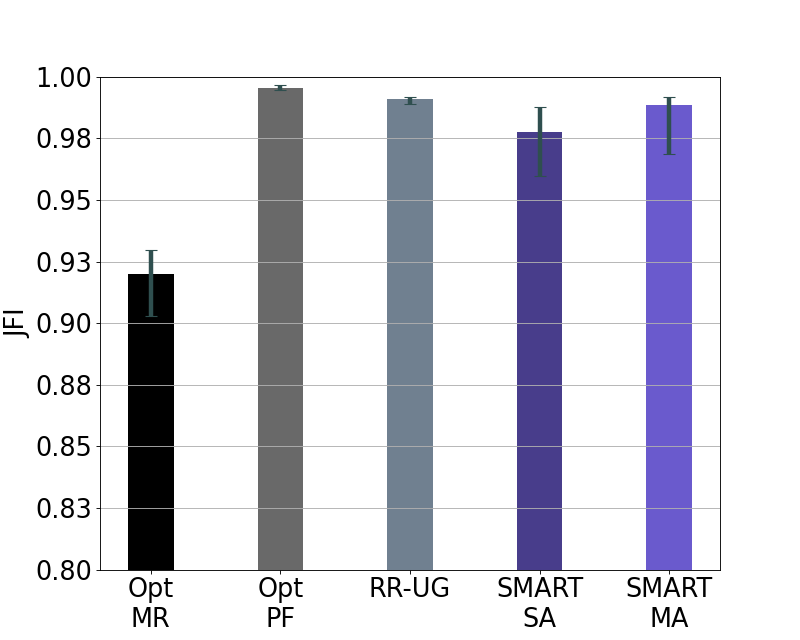}
      \caption{}
      \label{fig:2rb_jfi_8_random}
    \end{subfigure}
        \begin{subfigure}[b]{0.24\textwidth}
    \centering
        \includegraphics[width=\textwidth]{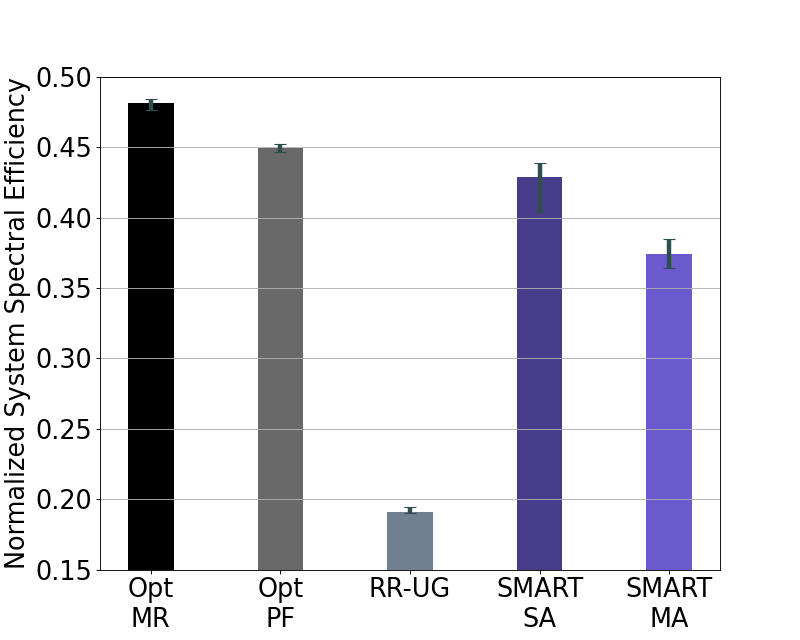}
        \caption{}
        \label{fig:2rb_se_8_mob}
    \end{subfigure}
    \hfill
    \begin{subfigure}[b]{0.24\textwidth}
    \centering
        \includegraphics[width=\textwidth]{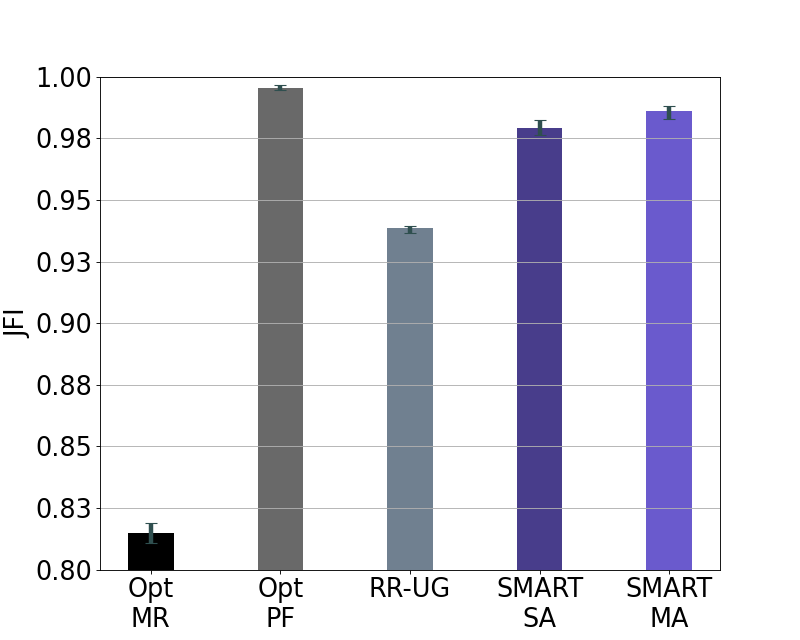}
        \caption{}
        \label{fig:2rb_jfi_8_mob}
    \end{subfigure}
    \caption{Spectral Efficiency and JFI comparison of \name{} and existing methods in $8\times 8$ network size and $N_{\mathrm{max}} = 4$ with 2 Resource Blocks in static random user topology (a) and (b), and user mobility scenario (c) and (d).} 
    \label{fig:2rb_se_jfi_8}
\end{figure}

%As discussed in~\ref{subsec:scalibility}, multi-agent DRL model suffers from convergence issues. To make it comparable, we first consider the configuration with 2 RBs, $M=8$, $L=8$, and $K_{max}=4$, random placement and mobility scenarios. Computation complexities of PF, RR-UG and MR under this configuration are also acceptable. Consequently, we also include them into our benchmarks. 
%But for Pointer Network, because its undesirable JFI performance in single RB experiment, it is destined to get worse for multiple blocks. To show our design's scalability, we also evaluate inter-user fairness performance of our scheduler in 100 RBs. In this configuration, PF and MR have too high computation complexity to be implemented and multi-agent DRL model have trouble with severe convergence problem. As a result, we only compare our design with RR-UG.
Here, we consider the multi-RB scenario and evaluate the performance of our model presented in~~\ref{subsec:scalibility}. As discussed, the multi-agent DRL models are generally difficult to converge. In fact, our \name{}-MA model only converged with 2 RBs ($B=2$) when $M=8$, $L=8$, and $N_{\mathrm{max}}=4$. Thus, we use this configuration to demonstrate the efficacy of \name{}-SA, with respect to \name{}-MA. Computational complexities of Opt-PF and Opt-MR were also acceptable in this configuration as presented in~\S\ref{sec:system_model}, and thus, we include them in the evaluation along with RR-UG. Since we showed the underwhelming performance of DDPG and PN in the single-RB case, we exclude them from this evaluation.
Fig.~\ref{fig:2rb_se_jfi_8} shows the experiment results for $B=2$. %\name{}-SA represents the single-agent DRL model and \name{}-MA denotes multi-agent DRL model. 
It is evident that \name{}-SA outperforms \name{}-MA on spectral efficiency but has a slightly lower JFI. The reason is that \name{}-SA tries to maximize spectral efficiency on each RB and sacrifices fairness as opposed to \name{}-MA which balances the two metrics across RBs. 
%On the opposite, \name{}-MA pays more attention to JFI and trades off spectral efficiency for it. 
\name{}-SA performs much better in terms of both JFI and spectral efficiency compared to RR-UG. 
%Based on the experiment results of 2 RBs, we can conclude that \name{}-SA has outstanding performance on spectral efficiency and the only concern is inter-user fairness when we extend it to tens or hundreds of RBs. Consequently, we exhibit JFI performance of $64\times 64$ network on 100 RBs in Fig.~\ref{fig:100_jfi}, which demonstrates that more RBs will not degrade JFI of \name{}-SA but also maintain desirable spectral efficiency. 
% Fig.~\ref{fig:100RB} 
For $B>2$, \name{}-MA, Opt-PF, and Opt-MR become infeasible. However, to demonstrate the performance of \name{}-SA, we evaluate it for $B=100$ with a $64\times 64$ network size and compare it with RR-UG. The evaluation results are shown in the simulation column of Table~\ref{tb3}.
For the results, it is evident that a large number of RBs will not degrade JFI in \name{}-SA while still maintaining desirable spectral efficiency. It also reaffirms our previous finding on the low performance of RR-UG in the mobility scenario.

\begin{table*}[t]\centering
% \caption{Spectral Efficiency and JFI comparison of \name{} and RR-UG in $64\times 7$ network size with 52 Resource Blocks in mobility LoS slow-speed topology, LoS high-speed topology and NLoS slow speed topology of real-world data}
\caption{Spectral Efficiency and JFI comparison of \name{} and RR-UG with multiple RBs in simulation discussed in~\S\ref{sec:scale} and with real-world data discussed in~\S\ref{subsec:realdataset}}
\begin{threeparttable}
\centering
\resizebox{\linewidth}{!}{
    {
    \begin{tabular}{c||cc|cc||cc|cc|cc} 
      \toprule
      \multirow{3}{*}{\textbf{Performance Metrics}} & \multicolumn{4}{c||}{\textbf{Simulation with $B=100$}} & \multicolumn{6}{c}{\textbf{Real-world Data with $B=52$}}\\
      \cmidrule{2-11}
      & \multicolumn{2}{c|}{\textbf{Random Placement}} & \multicolumn{2}{c||}{\textbf{Mobility Scenario}} & \multicolumn{2}{c|}{\textbf{LoS Slow-speed}} & \multicolumn{2}{c|}{\textbf{LoS High-speed}}& \multicolumn{2}{c}{\textbf{NLoS Slow-speed}} \\
      \cmidrule{2-11}
      & \textbf{\name{}-SA} & \textbf{RR-UG} &  \textbf{\name{}-SA} & \textbf{RR-UG} & \textbf{\name{}-SA} & \textbf{RR-UG}&  \textbf{\name{}-SA} & \textbf{RR-UG} & \textbf{\name{}-SA} & \textbf{RR-UG} \\
      \midrule
      Normalized System Spectral Efficiency & 0.500 & 0.254 & 0.400 & 0.063 & 0.713 & 0.662 & 0.670 & 0.584 & 0.488 & 0.481 \\
      \midrule
      JFI & 0.977 & 0.940 & 0.950 & 0.696 & 0.996 & 0.952 & 0.995 & 0.951 & 0.986 & 0.980 \\
     \bottomrule 
    \end{tabular}
    }
}
\end{threeparttable}
\label{tb3}
\end{table*}

\subsubsection{Real-World Data Evaluation}
\label{subsec:realdataset}
\begin{figure}[t]
      \centering
      \includegraphics[width=0.35\textwidth]{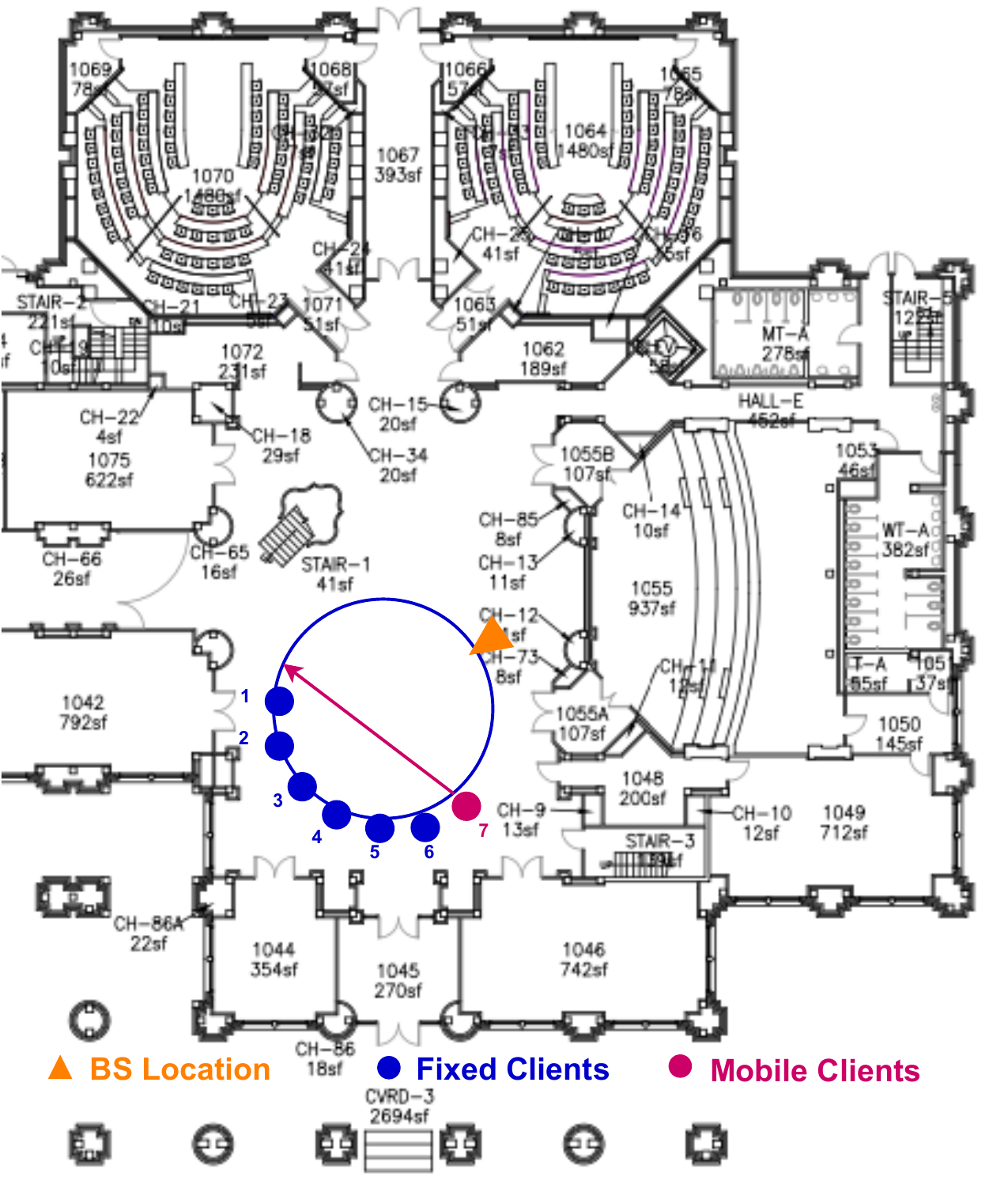}
      \caption{Topology of the real-world indoor experiment.} 
      \label{fig:exp_diag}
\end{figure}

To evaluate our proposed scheduler in real-world environments, we conducted a massive MIMO channel measurement experiment in an indoor setting on the Rice University campus. We used a 64-antenna RENEW~\cite{doost2018asilomar} software-defined massive MIMO base station and seven software-defined clients in a large open area inside a building hall. We fixed six of the clients in a circle, $15\mathrm{m}$ away from the base station. The seventh node was placed on a robot where we moved the robot across the hall starting from the location of the first client to the last. A drawing of the BS and client placements are shown in Fig.~\ref{fig:exp_diag}. We moved the robot along the path with different speeds, i.e. with $0.5 \mathrm{m/s}$, $1 \mathrm{m/s}$, and $2 \mathrm{m/s}$. The mobile node's antenna was facing the base station in all the experiments (LoS channel). We repeated the experiments to measure both LoS and NLoS channels for the fixed clients. 
In each measurement, we transmitted time-orthogonal uplink pilots from all clients to the BS. The uplink pilots were based on the 802.11 LTS OFDM signal, which contains 52 non-zero subcarriers. We consider each subcarrier as an RB in our evaluation, i.e. $B=52$.
Based on the collected real-world dataset, we train and evaluate the performance of \name{} in the $64\times 7$ MIMO configuration with 52 RBs in a slow-speed mobility scenario.

Using these datasets, we evaluate the performance of \name{}. Due to convergence issues and excessive computational complexity of other schedulers for $B>2$ as discussed in~\S\ref{sec:system_model}, we are only comparing \name{}-SA with RR-UG. The results, listed in Table~\ref{tb3},
%\textcolor{brown}{In our collected real-world dataset, we only have one mobile user. In this scenarios, in most cases, there are just one or two user groups including low inter-correlated users. For RR-UG scheduling strategy in mobility scenario as discussed in~\ref{subsec:bench}, it schedules seven clients in one TTI or utilizes two TTIs to schedule all user groups. It makes user grouping of RR-UG is not outdated in most TTIs so that RR-UG is a competitive benchmark to compare in real-world dataset. Due to the convergence issue and excessive computational complexity of other schedulers discussed in ~\S\ref{sec:scale}, for scheduling in 52 RBs, we are only compare \name{}-SA with RR-UG.}
% As shown in Fig.~\ref{fig:se_real_slow} and~\ref{fig:jfi_real_slow}, 
show that RR-UG underperforms \name{}-SA in both spectral efficiency and JFI. More importantly, \name{}-SA is capable of achieving near-optimal (i.e. about 0.996) JFI, which demonstrates the effectiveness of \name{}-SA when applied to multiple RBs. 
%\textcolor{red}{We should argue that the performance discrepancy will be much larger when there are mobile users.}
However, we can anticipate that RR-UG performance will get worse as the number of mobile users increases, which is consistent with the results of mobility scenarios in medium and real network size experiments.
By running Algorithm~\ref{alg1} on the datasets, we observe just one or two user groups in most TTIs. Thus, RR-UG schedules all seven clients in one or sometimes two TTIs. Therefore, RR-UG is rather competitive as \name{}-SA here.
For the purpose of showing the generality of our model, we use the model trained on the LoS slow-speed dataset and test it in the LoS high-speed mobility. 
% Fig.~\ref{fig:se_real_high} and~\ref{fig:jfi_real_high} 
The results in~Table~\ref{tb3} demonstrate the adaptability of \name{}-SA to different mobility scenarios. Compared with the slow-speed mobility scenario, it is obvious that the performance gap between \name{} and RR-UG in the high-speed scenario is larger. This is because a high speed makes channel condition and inter-user channel correlation vary more quickly than the slow speed. Faster varying inter-user channel correlation results in quicker variations of user grouping, which makes it challenging for RR-UG to adapt fast enough. However, \name{} is capable of dealing with this rapid change. 
For comprehensiveness, we also test the trained model on NLoS slow-speed topology. 
% Fig.~\ref{fig:se_real_nlos} and~\ref{fig:jfi_real_nlos} 
The results in~Table~\ref{tb3} show \name{}-SA's superiority over RR-UG and its generality in real-world data, albeit not as good as it is in LoS high-speed.

\subsubsection{Computational Complexity}
\label{sec:complexity}
We measure average wall-clock time per TTI for all the schedulers discussed in~\S\ref{sec:peformance}. For comparison fairness, we run all implementations on a single CPU core on the NVIDIA DGX server. The runtime values are listed in Table~\ref{tb4} for three network sizes considered in \S\ref{sec:peformance}. The results show the runtimes of the schedulers are widely different and they also vary with the network size. 
For Opt-MR and Opt-PF, the runtime increases exponentially with the network size and thus is not listed for network sizes beyond $16\times 16$. Even though Approx-PF is feasible in real-world size networks with much less complexity than Opt-PF, it still takes about $20\times$ times longer than ~\name{} to execute. Regarding other schedulers, the runtime seems to increase linearly. Both DDPG and \name{}-Vanilla show similar results. Comparing \name{} and \name{}-Vanilla results show that using user grouping labels instead of the raw channel matrix reduces the runtime of the model up to 50\%. Tuning hyper-parameters to achieve the best performance for both \name{} and \name{}-Vanilla, \name{} has 3 fewer hidden layers and half the number of neurons in each layer to remain on par with the performance of \name{}-Vanilla. However, user grouping requires only an additional 3.5 ms in $64\times 64$ network size, a negligible portion of the total runtime.
The runtime for PN is about 1.6x and 4x running time of \name{}-vanilla in $16\times 16$ and $64\times 64$, respectively. This is due to the fact that pointer networks are auto-regressive and make decisions sequentially and thus have slow inference. %This result also validates the conclusion in~\ref{subsec:bench}: Pointer Network is not desirable for real-time decision-making tasks. 
% \name{}-UM has slightly higher running time than \name{}-UG which shows the proposed scheduler can handle the additional complexity for finding the optimal modulation well. 
RR-UG shows the smallest runtime among all, but it is not as spectrally efficient as \name{}, especially in mobility scenarios.

\begin{table}[htp!]\centering
\caption{Wall-clock time in seconds per TTI}
\begin{threeparttable}
\centering
\resizebox{\linewidth}{!}{
    {
    \begin{tabular}{c||cccccccc} 
      \toprule
      \multirow{2}{*}{\textbf{System Configuration}} & \multicolumn{8}{c}{\textbf{Scheduler}} \\
      \cmidrule{2-9}
      & \textbf{Opt-MR} & \textbf{Opt-PF} & \textbf{Approx-PF} &  \textbf{RR-UG} & \textbf{DDPG} & \textbf{PN} & \textbf{\name-Vanilla}  & \textbf{\name} \\
      \midrule
    %   \textbf{8\_2\_4 (Static, \pmb{$\alpha:\beta = 1:1$})} & 6876 & 8107 & 10740 & 0.011 & 0.014 & 0.019\\
    %   \midrule
      $16\times 16$ & 0.15 & 0.21 & - &0.0013 & 0.034 & 0.059 & 0.036 & 0.024\\
      \midrule
     $64\times 64$ & - & - & 0.604 & 0.0043 & 0.058 & 0.235 & 0.057 & 0.030 \\
      \midrule
      $128\times 128$ & - & - & - & - & - & - & - & 0.071 \\
     \bottomrule 
    \end{tabular}
    }
}
\end{threeparttable}
\label{tb4}
\end{table}

\subsection{Discussion and Future Work}

\begin{figure}[t]
    \centering
    \begin{subfigure}[b]{0.24\textwidth}
    \centering
      \includegraphics[width=\textwidth]{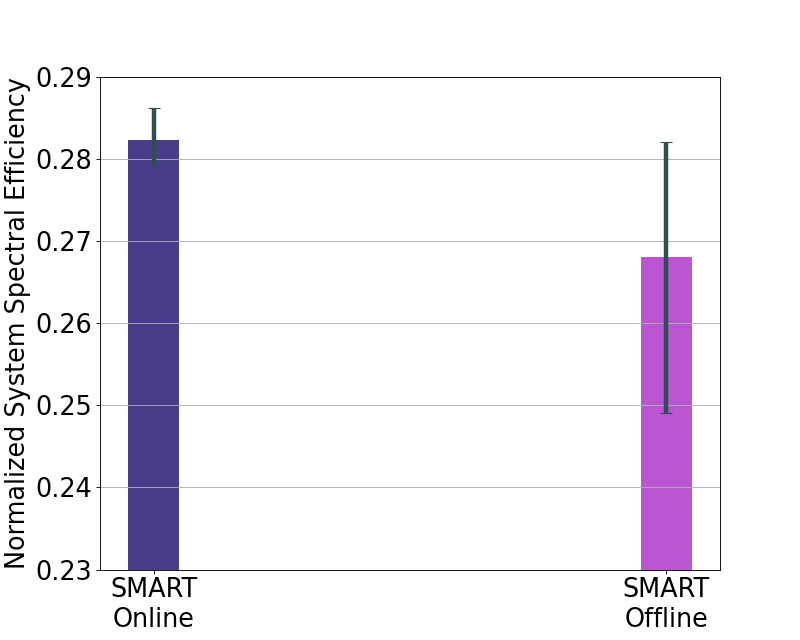}
      \caption{}
      \label{fig:se_64_mob_online}
    \end{subfigure}
    \hfill
    \begin{subfigure}[b]{0.24\textwidth}
    \centering
    \includegraphics[width=\textwidth]{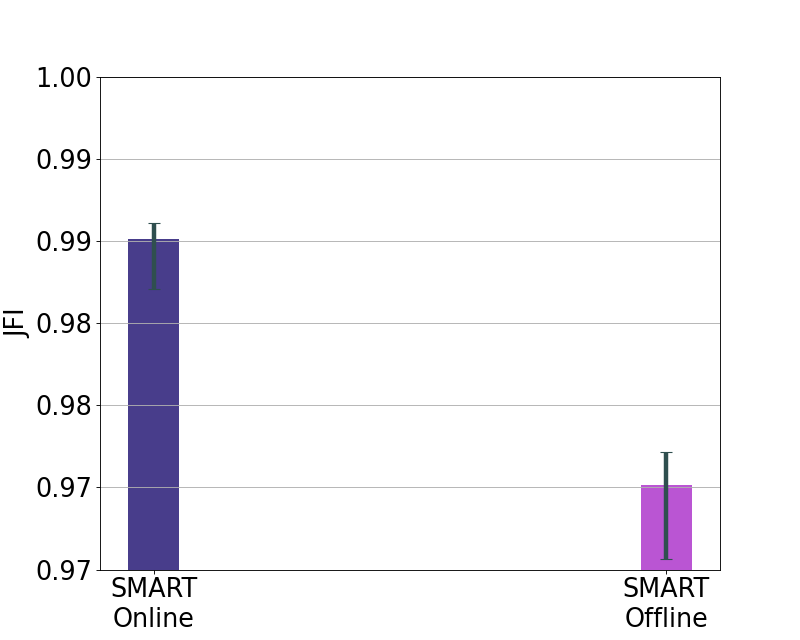}
      \caption{}
      \label{fig:jfi_64_mob_online}
    \end{subfigure}
    \caption{Evaluation of \name{} with and without a model ((online vs. offline) update in user mobility scenario and $64\times 64$ network size.} 
    \label{fig:offline}
\end{figure}

The results presented earlier offer good insights into the performance and computational complexity of the proposed \name{} scheduler with respect to the existing methods. However, an important question is whether \name{} can be deployed to operate in time-stringent 5G-NR systems. For a realistic network size, Table~\ref{tb4} shows \name{} takes as much as 30 $\mathrm{ms}$ to run an iteration, $30\times$ longer than one TTI in the least time-stringent mode of 5G-NR~\cite{chen2021mcore}. This may seem problematic for the adoption of \name{}. To investigate this, we run an experiment in a mobility scenario. We first train \name{} offline as before and test the trained model on the testing dataset without online updates to the model. We compare the spectral efficiency results for the offline trained model with the previously presented results that include the online updates. The results are shown in Fig.~\ref{fig:offline}. We observe that, even when we use the offline trained model with no online updates, the performance is remarkably close to when the model is continuously updated. The performance can get even closer when we do updates every few tens of TTIs. This finding means that we can only look into the inference time of the model as the scheduling decision time. For $16\times 16$ and $64\times 64$ network sizes, the inference times for \name{} are 5.4 and 8.7 $\mathrm{ms}$. Running the model on a single GPU core on the NVIDIA DGX A100 server reduces the inference time values to 1.2 and 1.6 $\mathrm{ms}$, respectively. The inference runtime values can be further reduced to sub-millisecond levels, as required in 5G-NR, by a more efficient implementation such as with CUDA~\cite{cuda} framework and parallelizing the DRL model on several GPU cores. More importantly, the reassuring performance of \name{}-SA, demonstrated in~\S\ref{sec:scale}, shows that we can get similar runtime values for 100s of RBs, as its architecture allows us to fully parallelize it on different GPU cores. 

Lastly, we have only considered saturated traffic for each user. A more generic design should consider the incoming traffic model as well as the quality of service (QoS) requirements, e.g. data rate and latency, for each user. Formulation of the scheduling problem and formally solving it using optimization techniques or heuristics-based approximation is a difficult task. We believe AI-based methods such as the one proposed in this paper provide a more promising avenue for solving the generic case if enough training data exists. We leave the design of a more comprehensive scheduler that considers parameters in the higher layers of the network such as traffic models and QoS constraints as future work.

%%%%%%%%% CONCLUSION
\section{Conclusion}
\label{sec:con}
In this paper, we presented \name{}, a resource scheduler for massive MIMO networks based on the soft actor-critic DRL model. We demonstrated the effectiveness of our scheduler in achieving both spectral efficiency as well as fairness very close to the optimal proportionally fair scheduler. We also showed that our model outperforms state-of-the-art massive MIMO schedulers in all scenarios, and particularly in mobility scenarios. We removed the need for raw channel matrices in training our DRL model by utilizing a user grouping algorithm based on the inter-user correlation matrix and, thus, we significantly reduced the complexity of our model. We also provided guidelines as to how our scheduling model can be deployed in time-stringent 5G-NR systems.

% \textcolor{red}{Please check capitalization in your references ... plenty of mimo or 5g, lans, 3gpp, etc. In fact, the first six references all of them have problems with the capitalization of some term, and many others across all references.}
%Virtualized radio access networks (vRANs) are the way base stations will be designed in the future. In this paper, we have presented SAC\_KNN, a vRAN solution that dynamically learns the optimal allocation of radio resources. Given a specific performance target, SAC\_KNN determines user selection and adopted modulation scheme to meet such target. To this end, SAC\_KNN builds on state-of-the-art deep reinforcement learning algorithm (SAC and KNN) to adapt different scales of user set. Our results shed light on the behavior of SAC\_KNN across different scenarios, MU-MIMO and massive MIMO, showing that SAC\_KNN is able to achieve the best system spectral efficiency while minimizing running time compared to other benchmarks. Moreover, inter-user fairness is close to optimal.

%%%%%%% -- PAPER CONTENT ENDS -- %%%%%%%%

% \section*{Acknowledgments}

% Lorem ipsum dolor sit amet, consectetur adipisicing elit, sed do eiusmod tempor incididunt ut labore et dolore magna aliqua. Ut enim ad minim veniam, quis nostrud exercitation ullamco laboris nisi ut aliquip ex ea commodo consequat. Duis aute irure dolor in reprehenderit in voluptate velit esse cillum dolore eu fugiat nulla pariatur. Excepteur sint occaecat cupidatat non proident, sunt in culpa qui officia deserunt mollit anim id est laborum.
% Generated by IEEEtran.bst, version: 1.14 (2015/08/26)

\bibliographystyle{IEEEtran} 
\begin{small}
\bibliography{refs}
\end{small}

\end{document}